\documentclass[onecolumn,showpacs,amsmath,nofootinbib,amssymb]{revtex4}

\usepackage{graphicx}
\usepackage{bm}

\newcommand{\Ta}[1]{\ensuremath{\Tilde{\alpha}_{#1}}}
\newcommand{\iTa}[1]{\ensuremath{\imath\Tilde{\alpha}_{#1}}}

\newcommand{\idTa}[1]{\ensuremath{\imath\delta\Tilde{\alpha}_{#1}}}

\newcommand{\Tb}{\ensuremath{\Tilde{\beta}}}

\newcommand{\Td}{\ensuremath{\Tilde{\delta}}}
\newcommand{\iTd}{\ensuremath{\imath\Tilde{\delta}}}
\newcommand{\TD}{\ensuremath{\Tilde{D}}}

\newcommand{\Te}{\ensuremath{\Tilde{\epsilon}}}
\newcommand{\iTe}{\ensuremath{\imath\Tilde{\epsilon}}}

\renewcommand{\TH}{\ensuremath{\Tilde{H}}}

\newcommand{\THee}{\ensuremath{\Tilde{H}_{ee}}}
\newcommand{\THbaree}{\ensuremath{\Tilde{\bar{H}}_{ee}}}
\newcommand{\THem}{\ensuremath{\Tilde{H}_{e\mu}}}
\newcommand{\THet}{\ensuremath{\Tilde{H}_{e\tau}}}
\newcommand{\THme}{\ensuremath{\Tilde{H}_{\mu e}}}
\newcommand{\THmm}{\ensuremath{\Tilde{H}_{\mu\mu}}}
\newcommand{\THbarmm}{\ensuremath{\Tilde{\bar{H}}_{\mu\mu}}}
\newcommand{\THmt}{\ensuremath{\Tilde{H}_{\mu\tau}}}
\newcommand{\THte}{\ensuremath{\Tilde{H}_{\tau e}}}
\newcommand{\THtm}{\ensuremath{\Tilde{H}_{\tau\mu}}}
\newcommand{\THtt}{\ensuremath{\Tilde{H}_{\tau\tau}}}
\newcommand{\THbartt}{\ensuremath{\Tilde{\bar{H}}_{\tau\tau}}}

\newcommand{\Hem}{\ensuremath{H_{e\mu}}}
\newcommand{\Hbarem}{\ensuremath{\bar{H}_{e\mu}}}
\newcommand{\Het}{\ensuremath{H_{e\tau}}}
\newcommand{\Hbaret}{\ensuremath{\bar{H}_{e\tau}}}
\newcommand{\Hme}{\ensuremath{H_{\mu e}}}
\newcommand{\Hbarme}{\ensuremath{\bar{H}_{\mu e}}}

\newcommand{\Hmt}{\ensuremath{H_{\mu\tau}}}
\newcommand{\Hbarmt}{\ensuremath{\bar{H}_{\mu\tau}}}
\newcommand{\Hte}{\ensuremath{H_{\tau e}}}
\newcommand{\Hbarte}{\ensuremath{\bar{H}_{\tau e}}}
\newcommand{\Htm}{\ensuremath{H_{\tau\mu}}}
\newcommand{\Hbartm}{\ensuremath{\bar{H}_{\tau\mu}}}

\newcommand{\dk}[1]{\ensuremath{\delta k_{#1}}}
\newcommand{\Tk}[1]{\ensuremath{\Tilde{k}_{#1}}}
\newcommand{\dTk}[1]{\ensuremath{\delta\Tilde{k}_{#1}}}

\newcommand{\dphi}[1]{\ensuremath{\delta\phi_{#1}}}

\newcommand{\Tc}[1]{\ensuremath{\Tilde{c}_{#1}}}
\newcommand{\Tce}{\ensuremath{\Tilde{c}_{\epsilon}}}

\newcommand{\Tnu}[1]{\ensuremath{\Tilde{\nu}_{#1}}}
\newcommand{\Tnubar}[1]{\ensuremath{\Tilde{\bar{\nu}}_{#1}}}

\newcommand{\TQ}{\ensuremath{\Tilde{Q}}}
\newcommand{\TQbar}{\ensuremath{\Tilde{\bar{Q}}}}
\newcommand{\TR}{\ensuremath{\Tilde{R}}}
\newcommand{\TRbar}{\ensuremath{\Tilde{\bar{R}}}}

\newcommand{\Ts}[1]{\ensuremath{\Tilde{s}_{#1}}}
\newcommand{\Tse}{\ensuremath{\Tilde{s}_{\epsilon}}}

\newcommand{\Tt}[1]{\ensuremath{\Tilde{\theta}_{#1}}}
\newcommand{\TT}{\ensuremath{\Tilde{T}}}

\newcommand{\TU}{\ensuremath{\Tilde{U}}}

\newcommand{\dV}[1]{\ensuremath{\delta V_{#1}}}

\newcommand{\x}[1]{\ensuremath{x_{#1}}}
\newcommand{\dx}[1]{\ensuremath{dx_{#1}}}

\begin{document}
\topmargin0.1in
\bibliographystyle{apsrev}
\title{Three Flavor Neutrino Oscillations in Matter: Flavor Diagonal Potentials, the Adiabatic Basis and the CP phase}

\author{James P. Kneller}
\email{Kneller@ipno.in2p3.fr}
\affiliation{School of Physics and Astronomy, University of Minnesota,
Minneapolis, Minnesota 55455,USA\\
Institut de Physique Nucl\'{e}aire Orsay, F-91406 Orsay cedex, France}

\author{Gail C. McLaughlin}
\email{Gail_McLaughlin@ncsu.edu}
\affiliation{Department of Physics, North Carolina State University, Raleigh,
North Carolina 27695-8202,USA}

\date{\today}

\pacs{14.60.Pq}

\begin{abstract}
We discuss the three neutrino flavor evolution problem with general,
flavor-diagonal, matter 
potentials and a fully parameterized mixing matrix that includes CP violation,
and
 derive expressions for the eigenvalues, mixing angles and phases.
We demonstrate that, in the limit that the mu and tau potentials are equal,
the eigenvalues and matter mixing angles $\tilde{\theta}_{12}$ and
$\tilde{\theta}_{13}$ 
are independent of the CP phase, although $\tilde{\theta}_{23}$ does have 
CP dependence. Since we are interested in developing
a framework that can be used for $S$ matrix calculations of neutrino flavor
transformation, 
it is useful to work in a basis that contains only  off-diagonal entries in the
Hamiltonian.
We derive the ``non-adiabaticity'' parameters that appear in the 
Hamiltonian in this basis. 
We then introduce the neutrino $S$ matrix, derive its evolution equation and the
integral solution.
We find that this new Hamiltonian, and therefore the $S$ matrix,  in the limit
that the mu and tau neutrino potentials are the same, is independent of both
$\tilde{\theta}_{23}$ and the CP violating phase. 
In this limit, any CP violation in the flavor basis can only be introduced via the rotation matrices, and so 
effects which derive from the CP phase are then straightforward to determine.
We show explicitly that the electron neutrino and electron antineutrino
survival probability is independent of the CP phase in this limit. Conversely,
if the CP phase is nonzero and mu and tau matter potentials are not equal, then
the electron neutrino survival probability cannot be independent of the CP phase.

\end{abstract}

\maketitle

%%%%%%%%%%%%%%%%%%%%%%%%%%%%%%%%%%%%%%%%%%%%%%%%%%%%%%%%%%%%%%%%%%%%%%%%%%%%%%%%
\newpage

\section{Introduction} \label{sec:intro}

Neutrino flavor transformation in vacuum
\cite{P1957,P1958,Maki:1962mu,Nakagawa:1963uw} and in matter
\cite{Wolfenstein1977,M&S1986} continues to be the subject of 
much attention both experimentally and theoretically. 
On the experimental side we have moved from a 
situation where little was known about
the mixing parameters  roughly a decade ago
to one where half of the mixing parameters are reasonably well known. 
The experimental status upon the mixing parameters is that 
$\delta m^{2}_{12}=8.0^{+0.4}_{-0.3}\;{\rm eV^{2}}$, 
$1.9\times10^{-3}\;{\rm eV^{2}}<|\delta m^{2}_{23}| < 3.0\times 10^{-3}\;{\rm
eV^{2}}$,
$\sin^{2}2\theta_{12}=0.86^{+0.03}_{-0.04}$, $\sin^{2}2\theta_{23}>0.92$, and 
$\sin^{2}2\theta_{13} < 0.19$ (at 90\%) \cite{PDG2006,2006PrPNP..57..742F}. 
The sign of $\delta m^{2}_{23}$, whether $\theta_{23}$ is less than or greater 
than $45^{\circ}$, and the extent of CP violation or the Majorana phases are not
currently known.

On the theoretical side the focus while 
 $\theta_{12}$ was completely unknown
was primarily upon solar neutrinos but after the results of SNO
\cite{2001PhRvL..87g1301A} 
development of neutrino flavor transformation theory  
has increasingly focused upon supernova neutrinos which display much 
richer phenomena. Two primary shifts in supernova neutrino flavor transformation
theory have occurred.
First, simplified and static profiles  have increasingly been abandoned, in
favor of more
realistic, dynamic and turbulent density profiles
\cite{Schirato:2002tg,Takahashi:2002yj,Fogli:2003dw,Tomas:2004gr,Fogli:2006xy,
K&M2006,Dasgupta:2005wn,Friedland:2006ta,Choubey:2007ga,2008PhRvD..77d5023K}.
Secondly, the implications of more complete descriptions of the potentials are
now being studied, such as 
neutrino self interactions 
\cite{Pantaleone:1992eq, Samuel:1993uw, Qian:1995ua,
Pastor:2002we, Balantekin:2004ug,
Sawyer:2005jk, Fuller:2005ae, Duan:2005cp, Duan:2006an,
Hannestad:2006nj, Balantekin:2006tg, Duan:2007mv, Raffelt:2007yz,
EstebanPretel:2007ec, Raffelt:2007cb, Raffelt:2007xt, Duan:2007fw,
Fogli:2007bk, Duan:2007bt, Duan:2007sh, EstebanPretel:2007yu, Dasgupta:2008cd,
EstebanPretel:2007yq, Dasgupta:2007ws, Duan:2008za, Dasgupta:2008my,
Sawyer:2008zs, Duan:2008eb, Chakraborty:2008zp,
Dasgupta:2008cu,EstebanPretel:2008ni,Sigl:2009cw}
and the ``matter mu-tau'' potential
\cite{Botella:1987aa,EstebanPretel:2007yq,Duan:2008za}.
These studies have uncovered new behaviors of the neutrinos, and in the process
in has become apparent that the calculational tools employed in the past have 
limited applicability to these newer situations. 

While some progress has been made in joining together these two new developments
 \cite{Gava:2009pj}, in this work we are interested primarily in the first. 
 An approximation that is often made in treating the evolution of the neutrino
wavefunction through more realistic density profiles is that 
the mixing of the neutrinos can be treated in a series of two flavor mixing
schemes even though 
there are three neutrino flavors. The computationally desirable reduction from
three to two 
flavors is based upon the observation that typically only two out of the three
neutrino states 
will be mixing with each other at a resonance and that resonances are widely
separated.  
But with the use of more realistic density profiles multiple, closely spaced,
resonances can occur 
and it has been shown \cite{K&M2006,Dasgupta:2005wn,2008PhRvD..77d5023K} that
finely grained energy dependent 
phase effects appear in two flavor mixing. This suggests that the phenomenon
also extends to three flavors, although
this has not yet been investigated. These effects can only be seen when the
phase information of
the wavefunctions is retained throughout the entire calculation, so most
approximation methods will miss them.
The full range of effects of the CP phase
\cite{Akhmedov:2002zj,Balantekin:2007es,2008arXiv0805.2717G} 
can also be hard to determine with approximation methods  
and these effects are purely three-flavor phenomenon. Finally, the small
difference between the matter potentials for the $\mu$ and $\tau$ flavors may also become important 
even in cases where the neutrino potentials are small, because this potential
difference produces a third matter resonance. 

As an alternative to the  usual integration of the neutrino wave functions, we
consider the $S$ matrix formulation of neutrino flavor transformation which was
shown to be computationally more efficient in the two flavor case
\cite{K&M2006}.  In this paper, we derive a convenient basis which can be used
with the $S$ matrix, which we call the ``adiabatic'' basis. We shall consider the
full, generalized three flavor mixing including 
all phases and all three matter potentials. In section \S\ref{sec:flavor}, we
derive the eigenvalues and discuss their behavior. In section \S\ref{sec:matter}
we present expressions for the matter mixing angles 
and phases and the Hamiltonian in the matter basis.
We then derive the three flavor Hamiltonian in the ``adiabatic basis''
\cite{K&M2006} 
- the basis where the Hamiltonian is completely off-diagonal - 
in section \S\ref{sec:adiabatic} and from it we discover the
expressions for the three-flavor non-adiabaticity parameters. 
We demonstrate how the $S$ matrix is found from this Hamiltonian in section
\S\ref{sec:to S}, its equation of motion and integral solution and discuss its
parameterization. We then discuss the two flavor approximation and observe some
identities of 
neutrino propagation that have implications for the CP phase.
Finally in section \S\ref{sec:conclusions} we present our conclusions.
\\
\\

%%%%%%%%%%%%%%%%%%%%%%%%%%%%%%%%%%%%%%%%%%%%%%%%%%%%%%%%%%%%%%%%%%%%%%%%%%%%%%%%

\section{The Flavor Basis} \label{sec:flavor}

In this section we write down the Hamiltonian which describes neutrino flavor
transformation in the flavor basis without approximations and determine its eigenvalues.

The $3 \times 3$ Hamiltonian $\TH^{(f)}$ in the 
flavor basis is composed of two terms: the rotated vacuum Hamiltonian $U\,K\,U^{\dag}$ 
and the matter induced potentials $V^{(f)}$'s which are diagonal in the flavor
basis if we ignore neutrino collective effects \cite{Pantaleone:1992xh}. 
The full Hamiltonian is 
\begin{equation}
\TH^{(f)} = U\,K\,U^{\dag}+ V^{(f)} \label{eq:Hf}
\end{equation}
where
%%
%%\begin{equation}
\begin{eqnarray}
K 
&=&
%%= 
\left(\begin{array}{lll} k_{1} & 0 & 0 \\ 0 & k_{2} & 0 \\ 0 & 0 & k_{3}
\end{array}\right) 
= E+\frac{1}{2E}\;\left(\begin{array}{lll} m_{1}^{2} & 0 & 0 \\ 0 & m_{2}^{2} &
0 \\ 0 & 0 & m_{3}^{2} \end{array}\right), 
\\
%%\end{equation}
%%
%%\begin{equation}
V^{(f)} 
&=&
%%= 
\left(\begin{array}{lll} V_{e} & 0 & 0 \\ 0 & V_{\mu} & 0 \\ 0 & 0
& V_{\tau} \end{array}\right),
\end{eqnarray}
%%\end{equation}
%%
and $E$ is the neutrino energy, $m_i$ the neutrino masses and $V_\alpha$ the 
matter potentials that are, possibly, functions of position.
We have included here three matter potentials for the neutrinos for generality. 
We could have removed one of these potentials, since we are free to add to the 
Hamiltonian an arbitrary multiple of the unit matrix 
including a term that is a function of position 
since the only effect of such a term is to introduce a phase. 
As a consequence of this property only the relative difference between
potentials (and eigenvalues)  
are important  for observable quantities, not the absolute values. 
However,  we choose not to impose this feature from the start, 
 and instead allow it to appear automatically as we proceed. For neutrinos, the
difference in the potentials $V_{e}$ and $V_\mu$ is the well known $\delta V_{e
\mu} \approx \sqrt{2}\,G_{F}\, (n_{e^{-}} -n_{e^{+}})$ \cite{Wolfenstein1977}
where $G_{F}$ is Fermi's constant and $n_{e}$ is the electron number density.
For antineutrinos, the potentials change sign. The potential difference between
$V_{\mu}$ and $V_{\tau}$ arises due to radiative corrections to neutral current
scattering and is smaller than $\dV{e\mu}$ and $\dV{e\tau}$ by a factor of $\sim
10^{-5}$ in typically encountered matter.  
Thus, the splitting is usually only important for neutrino propagation through supernova
profiles at high density.
 
In equation (\ref{eq:Hf}) $U$ is a unitary matrix that relates the flavor and
mass bases. 
The mixing matrix $U$ has nine elements but the unitary conditions mean that
four elements may be expressed in terms of the remaining five after specifying
the phase of the determinant.   
The unitary conditions also place two further restrictions upon the magnitudes
of the five independent elements 
by establishing two relationships between them. Thus, in general, $U$ is
parameterized by three magnitudes and six phases. (See e.g. Ref. 
\cite{Bronzan:1988wa} for discussion of the construction of an SU(3) matrix.)
It is traditional to select
the independent elements to be those in the first row and last column but other 
choices are also valid.
The three independent magnitudes, which must all be smaller than unity, may be
expressed in terms of three mixing angles $\theta_{12}$, $\theta_{13}$ and
$\theta_{23}$. 
The form for the matrix $U$ that we use for neutrinos is 
\begin{widetext}
\begin{eqnarray}
U & = & \left(\begin{array}{lll} 1 & 0 & 0\\ 0 & e^{\imath\delta} & 0\\0 & 0 & e^{\imath(\beta+\delta)}
\end{array}\right)
\left(\begin{array}{lll} c_{12}c_{13} & s_{12}c_{13} & s_{13}
\\ 
-s_{12}c_{23}\,e^{\imath\epsilon}-c_{12}s_{13}s_{23}
& c_{12}c_{23}\,e^{\imath\epsilon}-s_{12}s_{13}s_{23}
& c_{13}s_{23} \\ 
s_{12}s_{23}\,e^{\imath\epsilon}-c_{12}s_{13}c_{23} &
-c_{12}s_{23}\,e^{\imath\epsilon}-s_{12}s_{13}c_{23}
& c_{13}c_{23} \end{array}\right) 
%%\nonumber \\ && \qquad 
\left(\begin{array}{lll} e^{-\imath\alpha_{1}} & 0 & 0\\ 0 &
e^{-\imath\alpha_{2}} & 0\\0 & 0 & e^{-\imath\alpha_{3}}
\end{array}\right) \label{eq:U}
\end{eqnarray}
\end{widetext}
where $c_{12} = \cos \theta_{12}$, $s_{12} = \sin \theta_{12}$ etc.
and $\alpha_{1}$, $\alpha_{2}$, $\alpha_{3}$, $\beta$, $\delta$ and $\epsilon$
are the six phases. For antineutrinos, instead 
of $U$ we take  $U^{\star}$ for the unitary transformation matrix.
Other than the three $\alpha$'s we have tried to make the expression for $U$ as
similar as possible to that found in \cite{PDG2006} but note that in this more
general parametrization the CP phase is not $\delta$, but is instead $\epsilon$.
The phases $\beta$ and $\delta$ 
may be absorbed by redefinitions of the neutrino fields in the standard model
Lagrangian showing that their absolute values are 
not observable. This degree of freedom can be used to recover the 
usual convention for CP by setting $\delta$=$-\epsilon$. Even though the values 
of these two phases cannot affect observables we will continue to keep them
because, for consistency, 
during the calculation we must keep track of their derivatives.
The mixing matrix $U$ in equation (\ref{eq:U}) can be written as the product of
six terms
\begin{widetext}
\begin{subequations}
\begin{eqnarray}
U & =& B(\beta,\delta)\,\Theta_{23}(\theta_{23})\,E(\epsilon)
%%\nonumber \\ &&\qquad
\Theta_{13}(\theta_{13})\,\Theta_{12}(\theta_{12})\,A(\alpha_{1},\alpha_{2},\alpha_{3}), \\
&=& 
\left(\begin{array}{lll} 1 & 0 & 0 \\ 0 & e^{\imath\delta} & 0 \\ 0 & 0 &
e^{\imath(\beta+\delta)} \end{array}\right)
\left(\begin{array}{lll} 1 & 0 & 0 \\ 0 & c_{23} & s_{23} \\ 0 & -s_{23} &
c_{23} \end{array}\right)
%%\nonumber \\ &&
\left(\begin{array}{lll} 1 & 0 & 0 \\ 0 & e^{\imath\epsilon} & 0 \\ 0 & 0 & 1
\end{array}\right)
\left(\begin{array}{lll} c_{13} & 0 & s_{13} \\ 0 & 1 & 0 \\ -s_{13} & 0 &
c_{13} \end{array}\right)
%%\nonumber \\ &&
\left(\begin{array}{lll} c_{12} & s_{12} & 0 \\ -s_{12} & c_{12} & 0 \\ 0 & 0 &
1 \end{array}\right)
\left(\begin{array}{lll} e^{-\imath\alpha_{1}} & 0 & 0 \\ 0 &
e^{-\imath\alpha_{2}} & 0 \\ 0 & 0 & e^{-\imath\alpha_{3}}
\end{array}\right).
\nonumber \\
\label{eq:Udecomposition}
\end{eqnarray} 
\end{subequations}
%%\end{widetext}
%%
With this unitary matrix we find that the elements of $\TH^{(f)}$ are, in
full,
%%
%%\begin{widetext}
\begin{subequations}
\begin{eqnarray}
\THee & = & V_{e} + (k_{1}c_{12}^{2} + k_{2}s_{12}^{2})\,c_{13}^{2} + k_{3}s_{13}^{2}, \label{eq:Hee} \\
\THem & = & e^{-\imath\delta}\,\left[(k_{3} - k_{1}c_{12}^{2} -
k_{2}s_{12}^{2})\,c_{13}s_{13}s_{23} 
%%\right. \nonumber \\ && \left.
-(k_{1} - k_{2})\,c_{12}s_{12}c_{13}c_{23}e^{-\imath\epsilon}\right], \label{eq:Hem} \\
\THet & = & e^{-\imath(\beta+\delta)}\left[ (k_{3} - k_{1}c_{12}^{2} - k_{2}s_{12}^{2})\,c_{13}s_{13}c_{23} 
%%\right. \nonumber \\ && \left.
+ (k_{1} -k_{2})\,c_{12}s_{12}c_{13}s_{23}\,e^{-\imath\epsilon} \right],\label{eq:Het} \\
\THmm & = & V_{\mu} +(k_{1}+ k_{2})\,c_{23}^{2}+k_{3}c_{13}^{2}s_{23}^{2} 
%%\nonumber \\ && 
+(k_{1}c_{12}^{2} + k_{2}s_{12}^{2})\,(s_{13}^{2}s_{23}^{2}-c_{23}^{2})
%%\nonumber \\ && 
+2\,(k_{1}-k_{2})\,c_{12}s_{12}s_{13}c_{23}s_{23}c_{\epsilon},\label{eq:Hmm} \\
\THmt & = & e^{-\imath\beta}\left[
(k_{3}-k_{1}s_{12}^{2}-k_{2}c_{12}^{2})\,c_{23}s_{23}
%%\right. \nonumber \\ && \left.
-(k_{3}-k_{1}c_{12}^{2}-k_{2}s_{12}^{2})\,s_{13}^{2}c_{23}s_{23} 
%%\right. \nonumber \\ && \left.
+(k_{1}-k_{2})\,c_{12}s_{12}s_{13}\,(c_{23}^{2}\,e^{\imath\epsilon}-s_{23}^{2}\,
e^{-\imath\epsilon}) \right], \nonumber \\ \label{eq:Hmt} \\
\THtt & = & V_{\tau} +(k_{1}+ k_{2})\,s_{23}^{2}+k_{3}c_{13}^{2}c_{23}^{2} 
%%\nonumber \\ && 
+(k_{1}c_{12}^{2} + k_{2}s_{12}^{2})\,(s_{13}^{2}c_{23}^{2}-s_{23}^{2})
%%\nonumber \\ && 
-2\,(k_{1}-k_{2})\,c_{12}s_{12}s_{13}c_{23}s_{23}c_{\epsilon} \label{eq:Htt}.
\end{eqnarray}
\end{subequations}
%%\end{widetext}
%% 
The remaining elements are given by $\THme = \THem^{\ast}$, $\THte = \THet^{\ast}$ 
and $\THtm = \THmt^{\ast}$ but, since $V^{(f)}$ is diagonal, all the off-diagonal elements 
are equal to the vaccum values i.e. $\THem = \Hem$ etc. Note also that $\alpha_{1}$, $\alpha_{2}$ and $\alpha_{3}$ do not appear in these
expressions, although the phases $\delta$, $\beta$ and the CP phase $\epsilon$ do appear. 
We also point out that $\THmm+\THtt$, $\THmm\,\THmm-|\Hmt|^{2}$ and $|\Hem|^{2}+|\Het|^{2}$ are all 
independent of $\epsilon$.

The characteristic polynomial for the eigenvalues $\Tk{1}$, $\Tk{2}$ and
$\Tk{3}$ of $\TH$ can be written as 
\begin{equation}
\Tk{i}^{3} -\TT\,\Tk{i}^{2} + \left(\frac{\TT^{2}}{3} +3\,\TQ\right)\,\Tk{i} -
\left(\frac{\TT^{3}}{27} +\TQ\,\TT +2\,\TR\right) = 0 \label{eq:characteristic}
\end{equation}
where $\Tk{i}$ is any one of the three eigenvalues and $\TT$, $\TQ$ and $\TR$ are
three functions equal to
%%
%%\begin{widetext}
\begin{subequations}
\begin{eqnarray}
\TT& = & \THee+\THmm+\THtt, \label{eq:T} \\
\TQ & = & -\frac{1}{18}\left[\left(\THee-\THmm\right)^{2}
+\left(\THee-\THtt\right)^{2} +\left(\THmm-\THtt\right)^{2} \right]
-\frac{1}{3}\left[\left|\Hem\right|^{2}+\left|\Het\right|^{2}
+\left|\Hmt\right|^{2}\right], \label{eq:Q}\\
\TR& = &
\frac{1}{54}\left(2\,\THtt-\THee-\THmm\right)\left(2\,
\THmm-\THee-\THtt\right)\left(2\,\THee-\THmm-\THtt\right)
+\frac{1}{2}\left(\Hem\,\Hmt\,\Hte +\Het\,\Htm\,\Hme\right) \nonumber \\
&& -\frac{1}{6}\left[ \left|\Hmt\right|^{2}\,\left(2\,\THee-\THmm-\THtt\right)
+\left|\Het\right|^{2}\,\left(2\,\THmm-\THee-\THtt\right)
+\left|\Hem\right|^{2}\,\left(2\,\THtt-\THee-\THmm\right) \right]. \label{eq:R}
\end{eqnarray}
\end{subequations}
\end{widetext}
All three functions $\TT(V_{e},V_{\mu},V_{\tau})$, $\TQ(V_{e},V_{\mu},V_{\tau})$ and
$\TR(V_{e},V_{\mu},V_{\tau})$ are
polynomials in $V_{e}$, $V_{\mu}$, and $V_{\tau}$ and $\TT$, $\TQ$ and $\TR$ at a
given 
$V_{e}$, $V_{\mu}$ and $V_{\tau}$ are independent of the basis. In full these
polynomials are
\begin{subequations}
\begin{eqnarray}
\TT& = & T + V_{e} + V_{\mu} + V_{\tau}, \label{eq:T(V)}
\\
%%\end{eqnarray}\begin{eqnarray}
\TQ& = & Q +2\,q_{e}\,V_{e} +2\,q_{\mu}\,V_{\mu} +2\,q_{\tau}\,V_{\tau} 
%%\nonumber \\ && \qquad 
- \frac{1}{18}\left[ \dV{e\mu}^{2} + \dV{\mu\tau}^{2} + \dV{\tau e}^{2} \right], \label{eq:Q(V)}
\\
%%\end{eqnarray}\begin{eqnarray}
\TR& = & R +r_{e}\,V_{e} +r_{\mu}\,V_{\mu} +r_{\tau}\,V_{\tau} -q_{e}\,(V_{e}^{2}+2\,V_{\mu}\,V_{\tau}) 
%%\nonumber \\ && 
-q_{\mu}\,(V_{\mu}^{2}+2\,V_{e}\,V_{\tau}) -q_{\tau}\,(V_{\tau}^{2}+2\,V_{e}\,V_{\mu})
\nonumber \\ && \qquad
+\frac{1}{54}(\dV{e\mu}+\dV{e\tau})(\dV{\mu e}+\dV{\mu\tau})(\dV{\tau
e}+\dV{\tau\mu}), 
%%\nonumber \\ && 
\label{eq:R(V)}
\end{eqnarray}
\end{subequations}
where $T$, $Q$ and $R$ are the values of $\TT$, $\TQ$ and $\TR$ in the
vacuum, 
$\dV{\alpha\beta}=V_{\alpha}-V_{\beta}$, $\dV{\alpha\beta}^2
=(V_{\alpha}-V_{\beta})^2$ and the six functions $q_{e}$, $q_{\mu}$, $q_{\tau}$
and $r_{e}$, $r_{\mu}$, $r_{\tau}$ are 
\begin{subequations}
\begin{eqnarray}
q_{e} & = & \frac{1}{18}\left(H_{\mu\mu}+H_{\tau\tau}-2\,H_{ee} \right),
\label{eq:qe}\\
q_{\mu} & = & \frac{1}{18}\left(H_{ee}+H_{\tau\tau}-2\,H_{\mu\mu} \right),
\label{eq:qmu}\\
q_{\tau} & = & \frac{1}{18}\left(H_{ee}+H_{\mu\mu}-2\,H_{\tau\tau} \right),
\label{eq:qtau}\\
r_{e} & = & \frac{1}{18}\left[\left(H_{ee}-H_{\mu\mu}\right)^{2} +
\left(H_{ee}-H_{\tau\tau}\right)^{2} 
%%\right. \nonumber \\ && \left.
- 2\,\left(H_{\mu\mu}-H_{\tau\tau}\right)^{2} 
%%\right. \nonumber \\ && \left.
+3\,\left(\left|\Hem\right|^{2}+\left|\Het\right|^{2}-2\,\left|\Hmt\right|^{2}
\right) \right],\label{eq:re} \\
r_{\mu} & = & \frac{1}{18}\left[\left(H_{\mu\mu}-H_{\tau\tau}\right)^{2} +
\left(H_{\mu\mu}-H_{ee}\right)^{2} 
%%\right. \nonumber \\ && \left.
- 2\,\left(H_{\tau\tau}-H_{ee}\right)^{2} 
%%\right. \nonumber \\ && \left.
+3\,\left(\left|\Hmt\right|^{2}+\left|\Hme\right|^{2}-2\,\left|\Hte\right|^{2}
\right) \right],\label{eq:rmu} \\
r_{\tau} & = & \frac{1}{18}\left[\left(H_{\tau\tau}-H_{ee}\right)^{2} +
\left(H_{\tau\tau}-H_{\mu\mu}\right)^{2} 
%%\right. \nonumber \\ && \left.
- 2\,\left(H_{ee}-H_{\mu\mu}\right)^{2}
%%\right. \nonumber \\ && \left.
+3\,\left(\left|\Hte\right|^{2}+\left|\Htm\right|^{2}-2\,\left|\Hem\right|^{2}
\right) \right].\label{eq:rtau}
\end{eqnarray}
\end{subequations}
In the above expressions, we use $H_{ee}$ etc. to refer to the first element of 
$H^{(f)} = U\,K\,U^{\dag}$, i.e the flavor basis Hamiltonian in vacuum. Since the quantities,
$\TT$, $\TQ$ and $\TR$ are independent of the basis and their vacuum values are most
easily expressed 
in the diagonal, mass basis. In this basis their values are
\begin{subequations}
\begin{eqnarray}
T & = & k_{1}+k_{2}+k_{3}, \label{eq:TV}\\
Q & = & -\frac{1}{18}\left[ \dk{12}^{2} + \dk{13}^{2} + \dk{23}^{2}
\right],\label{eq:QV}\\
R & = & \frac{1}{54}\left[
(\dk{12}+\dk{13})\,(\dk{21}+\dk{23})\,(\dk{31}+\dk{32})\right],
%%\nonumber \\&& 
\label{eq:RV}
\end{eqnarray}
\end{subequations}
where $\dk{ij} = k_{i} - k_{j}$ and $\dk{ij}^2 = (k_{i} - k_{j})^2$.
The discriminant $\TD=\TQ^{3}+\TR^{2}$ is negative definite 
consequently all three eigenvalues are real. In terms of $\TT$, $\TQ$ and $\TR$ the 
eigenvalues $\Tk{1}$, $\Tk{2}$ and $\Tk{3}$ are \cite{Betal80}:
\begin{equation}
\Tk{i} = \frac{\TT}{3} + 2\,\sqrt{-\TQ}\,\cos\left( \frac{\omega + \omega_{i}}{3}
\right), \label{eq:k} \\
\end{equation}
where $\cos \omega = \TR /\sqrt{-\TQ^{3}}$. The three angles 
$\omega_{1}$, $\omega_{2}$ and $\omega_{3}$ are chosen to 
ensure that each eigenvalue 
takes on its appropriate value in vacuum, i.e $\Tk{1}$ becomes $k_{1}$ in the
vacuum and similarly $\Tk{2} \rightarrow k_{2}$ and 
$\Tk{3} \rightarrow k_{3}$. As the orderings of the vacuum values depend on the
hierarchy, 
there are two possible choices of values for these angles 
$\omega_{1}$, $\omega_{2}$ and $\omega_{3}$.  In the case of the normal
hierarchy (NH) where 
$\Tk{1} < \Tk{2} < \Tk{3}$, the angles must be $\omega_{1}=2\pi$,
$\omega_{2}=4\pi$ and $\omega_{3}=0$, while
 in the case of the inverted  hierarchy (IH) where $\Tk{3} < \Tk{1} < \Tk{2}$,
the angles must be $\omega_{1}=4\pi$, 
$\omega_{2}=0$ and $\omega_{3}=2\pi$.  Antineutrinos have the same vacuum
eigenvalues as neutrinos, so these
angles are the same for both neutrinos and antineutrinos. 

In the limit where $V_\mu$=$V_\tau$=0 we find that the characteristic
polynomial, equation (\ref{eq:characteristic}), is independent of the CP phase. It is clear
that the vaccum values of $T$, $Q$, and $R$, \ref{eq:TV}-\ref{eq:RV}, are independent of the CP phase, but a
careful inspection of equations (\ref{eq:T(V)}) - (\ref{eq:R(V)}) using previously established 
identites between the elements of the Hamiltonian shows that the $\epsilon$
dependence drops out here as well. This implies the eigenvalues are also independent of $\epsilon$.
\begin{figure*}
\includegraphics[width=5in]
{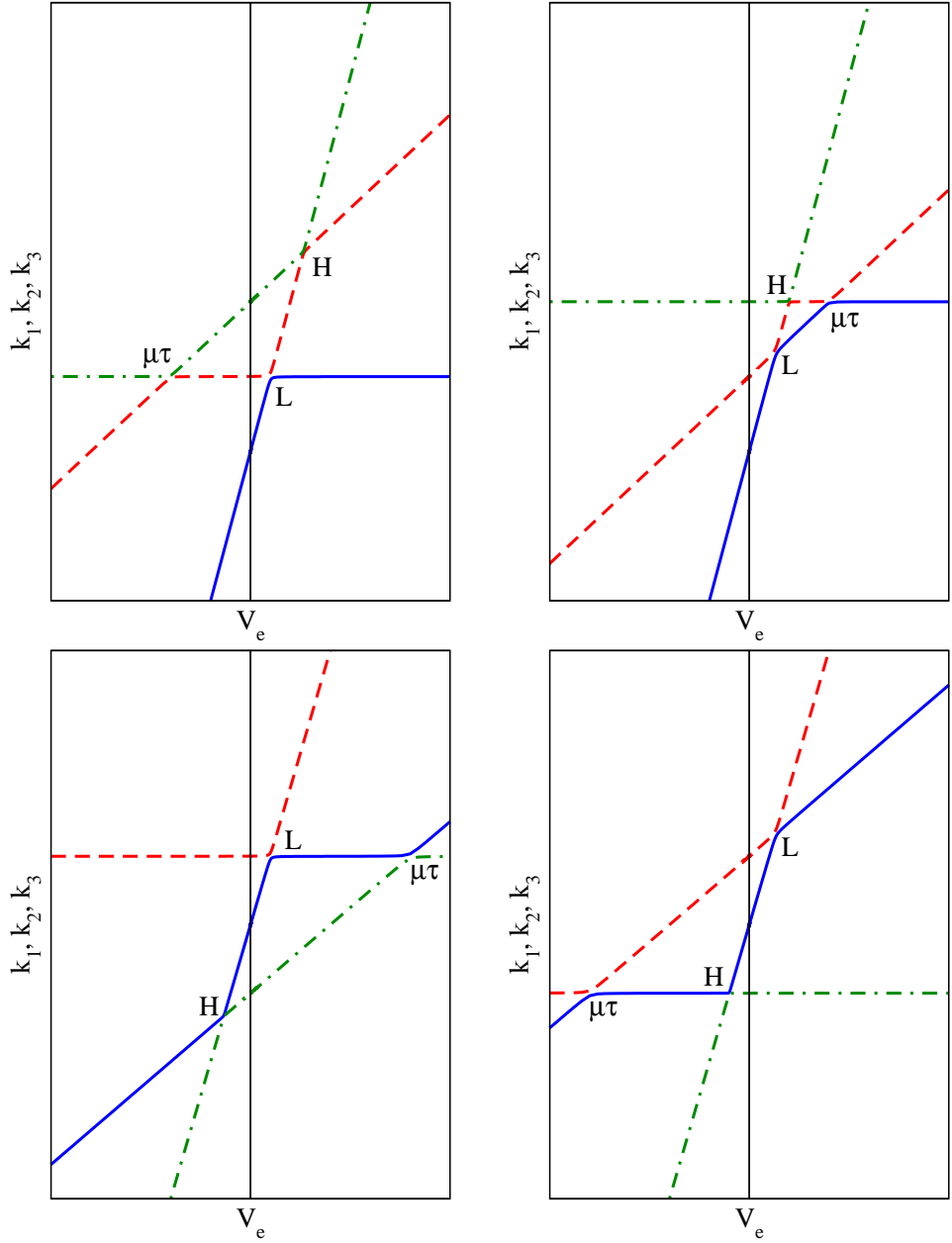}
\caption{The four possibilities for ordering and evolution of the eigenvalues
$\Tk{1}$, $\Tk{2}$ and $\Tk{3}$. 
The potential $V_{\mu}$ is set to zero while $V_{\tau} \propto V_{e}$ but
$|V_{e}| > |V_{\tau}|$. 
The solid vertical line is located at $V_{e}=0$. The vacuum mixing angles
$\theta_{12}$ and 
$\theta_{13}$ were set to small values so that the eigenvalues closely followed
the potentials.
In all four panels the solid line is $\Tk{1}$, the short dashed line is $\Tk{2}$
and the long dashed line is 
$\Tk{3}$. In the top two panels the hierarchy is normal, and in the bottom pair
the hierarchy is inverted (IH). 
For the left pair of panels the vacuum mixing angle $\theta_{23}$ obeys
$\theta_{23} < \pi/4$, for the right pair $\theta_{23} > \pi/4$. The units of
either axis are linear but arbitrary. \label{fig:k1k2k3}}
\end{figure*}

In addition to the hierarchy, the behavior of the matter eigenstates is strongly
influenced by 
whether $\theta_{23} > \pi/4$ or $\theta_{23} < \pi/4$. Thus, when considering
all the matter potentials,  
our present knowledge of the neutrino mixing parameters allows 
four possibilities for the general evolution of $\Tk{1}$, $\Tk{2}$ and $\Tk{3}$ 
\cite{EstebanPretel:2007yq} with $V$. 
These are shown in figure (\ref{fig:k1k2k3}). The solid vertical line in the
middle of each panel indicates vacuum with $V_e = 0$, so the left portion of
each panel corresponds to antineutrinos and the right portion to neutrinos. 
Note that in the figure, the axes are scaled linearly, not logarithmically. 
To make the figure we set both $\theta_{12}$ and $\theta_{13}$ to 
be small and included a non-zero potential for $V_{\tau}$ proportional to, but
smaller in magnitude than, $V_{e}$ and set $V_{\mu} = 0$. In reality,
$\theta_{12} \approx 34$ degrees, but the behavior of the eigenstates is still
qualitatively similar to what is shown in the figure. Also  $V_{\mu}$ is not
zero, but in order to best illustrate the general behavior of the eigenvalues in
a figure, we have used the freedom of adding to the Hamiltonian a multiple of
the unit matrix.

When the mixing angles are small each eigenvalue tends to track one of the potentials. At
``resonances'' they switch which potential they follow therefore ensuring that the
ordering of the eigenvalues is invariant. With 
three potentials there are three resonances commonly referred to as the ``L'', ``H''
and the ``$\mu\tau$''. The L resonance, L standing for ``low density'', occurs at
the smallest positive value of $V_{e}$ and, because $\delta m^{2}_{21}$ is known
to be 
positive, the two states that mix are always the neutrino states
$\Tilde{\nu}_{1}$ and $\Tilde{\nu}_{2}$. The H resonance, H standing for ``high
density'' occurs at the intermediate value of $|V_{e}|$. Which two states mix
depends upon the hierarchy: for a normal 
hierarchy (NH) it is states $\Tnu{2}$ and $\Tnu{3}$, for an
inverted hierarchy (IH) it is the antineutrinos states $\Tnubar{1}$
and $\Tnubar{3}$. The two states that mix at the $\mu\tau$ resonance, which occurs at the highest
value of $|V_{e}|$ 
depends both upon the hierarchy and whether $\theta_{23} < \pi/4$ or
$\theta_{23} > \pi/4$. For a NH and $\theta_{23} < \pi/4$ it is states
$\Tnu{1}$ and $\Tnu{2}$ but if $\theta_{23} > \pi/4$ then it is
states $\Tnubar{2}$ and $\Tnubar{3}$, in an IH states
$\Tnubar{1}$ and $\Tnubar{2}$ mix when $\theta_{23} <
\pi/4$ but $\Tnu{1}$ and $\Tnu{3}$ when $\theta_{23} > \pi/4$.
Out of the four possibilities the only one which causes two eigenvalues to mix
twice, when considering 
only a monotonically decreasing density profile,  
is the scheme of a NH and $\theta_{23} < \pi/4$ because one sees that states
$\Tnu{1}$ and $\Tnu{2}$ mix once at the $\mu\tau$ resonance and
again at the L resonance. This double mixing of $\Tnu{1}$ and $\Tnu{2}$ 
raises the possibility that interference effects might appear even in monotonic profiles. 
We will examine the consequences of this in future
work, but now we use the results of this section to compute the Hamiltonian in
the matter basis for use with the $S$ matrix.

%%%%%%%%%%%%%%%%%%%%%%%%%%%%%%%%%%%%%%%%%%%%%%%%%%%%%%%%%%%%%%%%%%%%%%%%
%%%%%%%%%%%%%%%%%%%%%%%%%%%%%%%%%%%%%%%%%%%%%%%%%%%%%%%%%%%%%%%%%%%%%%%%
%%%%%%%%%%%%%%%%%%%%%%%%%%%%%%%%%%%%%%%%%%%%%%%%%%%%%%%%%%%%%%%%%%%%%%%%

\section{The Matter Basis}\label{sec:matter}

In this section, while keeping track of all the angles, phases and their
derivatives, we write down the Hamiltonian in the matter basis.  We then make a
few remarks about the large density limit.

With the eigenvalues determined we can begin by deducing the angles $\Tt{12}$,
$\Tt{13}$, $\Tt{23}$ and phases $\Tb$, $\Td$ and $\Te$ in the unitary
transformation $\TU$ that relates $\TH^{(f)}$ to the matrix where $\Tk{1}$,
$\Tk{2}$ and $\Tk{3}$ appear on the diagonal i.e
\begin{equation}
\TU^{\dag}\,\TH^{(f)}\,\TU = \Tilde{K} =\left(\begin{array}{lll} \Tk{1} & 0
& 0 \\ 0 & \Tk{2} & 0 \\ 0 & 0 & \Tk{3} \end{array}\right).
\end{equation}
After making the change of basis we find that the Schrodinger equation is 

\begin{subequations}
\begin{eqnarray}
\imath \frac{d\psi^{(m)}}{dx} & = & \left( \Tilde{K} -\imath
\TU^{\dag}\frac{d\TU}{dx} \right)\;\psi^{(m)}.
\\
%%\end{eqnarray}\begin{eqnarray}
& = & \TH^{(m)}\;\psi^{(m)}.
\end{eqnarray}
\end{subequations}
The term $\TU^{\dag}\,d\TU/dx$ appears because the eigenvalues are functions of
position which requires 
that $\TU$ also be a function of position. Obviously if $\TU^{\dag}\,d\TU/dx =
0$ then the matter basis Hamiltonian would be diagonal but this can only occur
if the density is constant otherwise, in general, $\TH^{(m)}$ has non-zero
off-diagonal elements that come from $\TU^{\dag}\,d\TU/dx$. In order to evaluate
$\TH^{(m)}$ we need to compute $\TU^{\dag}\,d\TU/dx$.

The matter mixing matrix $\TU$ is parameterized in exactly the same way as $U$
in equation (\ref{eq:U}) i.e. as
\begin{widetext}
\begin{eqnarray}
\TU & = & \left(\begin{array}{lll} 1 & 0 & 0\\ 0 & e^{\iTd} & 0\\0 & 0 & e^{\imath(\Tb+\Td)}
\end{array}\right)\,
\left(\begin{array}{lll} 
\Tc{12}\Tc{13} & \Ts{12}\Tc{13} & \Ts{13}\,e \\ 
-\Ts{12}\Tc{23}\,e^{\iTe}-\Tc{12}\Ts{13}\Ts{23} &
\Tc{12}\Tc{23}\,e^{\iTe}-\Ts{12}\Ts{13}\Ts{23} &
\Tc{13}\Ts{23} \\ 
\Ts{12}\Ts{23}\,e^{\iTe}-\Tc{12}\Ts{13}\Tc{23} &
-\Tc{12}\Ts{23}\,e^{\iTe}-\Ts{12}\Ts{13}\Tc{23} &
\Tc{13}\Tc{23} \\
\end{array}\right) 
%%\nonumber \\ && \qquad
\left(\begin{array}{lll} e^{-\iTa{1}} & 0 & 0\\ 0 & e^{-\iTa{2}}
& 0\\0 & 0 & e^{-\iTa{3}}  \end{array}\right) \label{eq:tildeU}
\end{eqnarray}
where $\Tc{ij} = \cos \Tt{ij}$ etc. After some 
rather lengthy algebra we find that the matter mixing angles $\Tt{12}$,
$\Tt{13}$ and $\Tt{23}$, and the matter phases $\Tb$, $\Td$ and $\Te$ can be
expressed in terms of the eigenvalues and the elements of the flavor basis
Hamiltonian as
\begin{subequations}
\begin{eqnarray}
\tan^{2} \Tt{12} & = & - \frac{\dTk{13}\,\left[(\THmm-\Tk{2})(\THtt-\Tk{2}) -
\left|\Hmt\right|^2\right] }{ \dTk{23}\,\left[(\THmm-\Tk{1})(\THtt-\Tk{1}) -
\left|\Hmt\right|^2\right]}, \label{eq:theta12}\\
\sin^{2} \Tt{13} & = & \frac{(\THmm-\Tk{3})\,(\THtt-\Tk{3})
-\left|\Hmt\right|^{2}}{\dTk{13} \,\dTk{23} },\label{eq:theta13}\\
\tan^{2} \Tt{23} & = & \left(
\frac{\Hmt\,\Hte-\Hme(\THtt-\Tk{3})}{\Htm\,\Hme-\Hte(\THmm-\Tk{3})}
\right)\left(
\frac{\Het\,\Htm-\Hem(\THtt-\Tk{3})}{\Hem\,\Hmt-\Het(\THmm-\Tk{3})}
\right), \label{eq:theta23} 
\\
\imath\,\tan\Tb  & = & \frac{ \Hte\,\Hem - \Hme\,\Het - (\Htm-\Hmt)(\THee
- \Tk{3}) }{ \Hte\,\Hem + \Hme\,\Het - (\Htm + \Hmt)\,(\THee - \Tk{3})},
\label{eq:beta}\\
\imath\,\tan\Td  & = & \frac{ \Hmt\,\Hte - \Het\,\Htm - (\Hme-\Hem)(\THtt
- \Tk{3}) }{ \Hmt\,\Hte + \Het\,\Htm - (\Hme + \Hem)\,(\THtt -
\Tk{3})},\label{eq:delta1}\\
\sin\Te & = &
\frac{\dk{12}\dk{13}\dk{23}c_{12}s_{12}c_{13}^{2}s_{13}c_{23}s_{23}}{\dTk{12}
\dTk{13}\dTk{23}\Tc{12}\Ts{12}\Tc{13}^{2}\Ts{13}\Tc{23}\Ts{23}}\,\sin\epsilon.
\label{eq:delta2}
\end{eqnarray}
\end{subequations}
\end{widetext}
These equations match those already written down in the literature in the
appropriate limits.
The last equation, (\ref{eq:delta2}), for $\Te$ is the Naumov \cite{Naumov92}
and Harrison \& Scott \cite{HS00} identity and comes from the fact that the
off-diagonal elements of $\TH$ given in equations (\ref{eq:Hem}), (\ref{eq:Het})
and (\ref{eq:Hmt}) are not functions of position. 
The Toshev \cite{T91} and Kimura, Takamura \& Yokomakura
\cite{2002PhRvD..66g3005K} identities are no longer valid when $\dV{\mu\tau}
\neq 0$. Other than a re-arrangement of the 
ordering of the eigenvalues and some notational changes the expressions for 
$\tan^{2} \Tt{12}$ and $\sin^{2} \Tt{13}$ are the same as those one may find in
Bellandi \emph{et al.} 
\cite{1997BrJPh..27..384B}; our expression for $\tan^{2} \Tt{23}$ differs
from Bellandi \emph{et al.} because we allow the possibility of $\epsilon \neq
0$ and do not set $\beta = 0$. 
When performing numerical calculations,
the expressions (\ref{eq:theta12}) - (\ref{eq:delta2}) become increasingly
difficult to use when the densities become large. In section \ref{sec:limit} we discuss this limit, 
how one can derive asymptotic expressions for equations (\ref{eq:theta12}) to
(\ref{eq:delta2}) and demonstrate why the numerical difficulties arise. 

The most straightforward way to use these equations is to first derive the
eigenvalues $\Tk{1},\Tk{2},\Tk{3}$
at each density and then determine the matter phases and angles. But since
these are trigonometric expressions, more than one value of the angles or phases can satisfy the identity. The matter
angles  $\Tt{12}$ and $\Tt{13}$, always remain the quadrant in which they are first defined, which removes any
ambiguity in using equations (\ref{eq:theta12}) and (\ref{eq:theta13}).
Furthermore, these two angles are independent of $\epsilon$ if $V_\mu=V_\tau$. This can be observed 
because we have already established that the eigenvalues and the 
expressions $\THmm+\THtt$, $\THmm\,\THtt-|\Hmt|^{2}$ are independent of $\epsilon$.
However $\Tt{23}$ can migrate with $V$ into the second quadrant if its vacuum
value is defined to lie in the first quadrant so we do have an ambiguity to resolve when using equation
(\ref{eq:theta23}) to find $\Tt{23}$. An examiniation of equation (\ref{eq:theta23}) shows that 
$\Tt{23}=\pi/2$ when 
\begin{equation}
\Tk{3}=\THmm-\frac{\Htm\,\Hme}{\Hte} \equiv
k_{90},\;\,\Tk{3}=\THmm-\frac{\Hem\,\Hmt}{\Het} \equiv
k_{90}^{\star}.\label{eq:k3star}
\end{equation}
In general $k_{90}$ is complex and so at no density can we obtain an eigenvalue $\Tk{3}$ that satisfies the condition and, therefore,
$\Tt{23}$ never wanders out from the first quadrant. But when $\Hem\,\Hmt\,\Hte$ is pure real (which requires the CP
violating phase $\epsilon$ to be zero or $\pi$) then $k_{90}$ is also pure real
and therefore there is a density at which $\Tt{23}$ approaches and then passes $\pi/2$. 
Thus we see that the evolution of $\Tt{23}$ with the potential depends upon $\epsilon$ in a way that $\Tt{12}$ and 
$\Tt{13}$ do not. 
If we have a case where $\epsilon$ is either zero or $\pi$ then by keeping track of the behavior of these denominators, equation (\ref{eq:k3star}), we can resolve the ambiguity and determine the correct quadrant for $\Tt{23}$. 
If we have set $\theta_{23}$ in the vacuum to be less than $\pi/2$  {\it and} $k_{3} < k_{90}$ then
$\Tt{23}$ must become greater than $\pi/2$ for $\Tk{3} > k_{90}$.  
If on the other hand we have set $\theta_{23}$ in the vacuum to be less than
$\pi/2$ \emph{but} $k_{3} > k_{90}$, then $\Tt{23}$ must remain in the first quadrant for all positive definite $V$. Similar
arguments can be made for the zeros of the numerator of equation (\ref{eq:theta23}) that become relevant in
the inverted hierarchy. One finds that the numerators vanish at the root $k_{0}$ and, again, $k_{0}$ is
generally complex except when $\epsilon$ is either zero or $\pi$. So, if $\theta_{23}$ in the
vacuum is greater than zero and $k_{3} > k_{0}$ then $\Tt{23} < 0$ if $\Tk{3} < k_{0}$.
The CP phase $\epsilon$ can range from zero to $\pi$, but since we can determine not only $\sin \Te$ 
but also $\cos \Te$ there is no ambiguity in the CP phase. In principle,
there are ambiguities in the phases $\beta$ and $\delta$, but in practice, only
their derivatives are used not their absolute values. Finally we note there are
no expressions for $\Ta{1}$, $\Ta{2}$ and $\Ta{3}$ so we are free to pick
anything for them \emph{including} functions of $x$. 
Finally, once the matter mixing angles and phases are found we note that, like $U$, the
mixing matrix $\TU$ can be written as the product of six terms
\begin{subequations}
\begin{eqnarray}
\TU & =&
B(\Tb,\Td)\,\Theta_{23}(\Tt{23})\,E(\Te)\,
%%\nonumber \\ && \qquad 
\Theta_{13}(\Tt{13})\,\Theta_{12}(\Tt{12})\,A(\Ta{1},\Ta{2},\Ta{3}), 
\\ &=& 
\left(\begin{array}{lll} 1 & 0 & 0 \\ 0 & e^{\iTd} & 0 \\ 0 & 0 &
e^{\imath(\Tb+\Td)} \end{array}\right)
\left(\begin{array}{lll} 1 & 0 & 0 \\ 0 & \Tc{23} & \Ts{23} \\ 0 & -\Ts{23} &
\Tc{23} \end{array}\right)
%%\nonumber \\ && 
\left(\begin{array}{lll} 1 & 0 & 0 \\ 0 & e^{\iTe} & 0 \\ 0 & 0 & 1
\end{array}\right)
\left(\begin{array}{lll} \Tc{13} & 0 & \Ts{13} \\ 0 & 1 & 0 \\ -\Ts{13} & 0 &
\Tc{13} \end{array}\right)
%%\nonumber \\ && 
\left(\begin{array}{lll} \Tc{12} & \Ts{12} & 0 \\ -\Ts{12} & \Tc{12} & 0 \\ 0 &
0 & 1 \end{array}\right)
\left(\begin{array}{lll} e^{-\iTa{1}} & 0 & 0 \\ 0 & e^{-\iTa{2}} & 0 \\ 0 & 0 &
e^{-\iTa{3}} \end{array}\right).
\nonumber \\&& 
\label{eq:tildeUdecomposition}
\end{eqnarray} 
\end{subequations}

Next we need to calculate the derivatives of these angles and phases. A
surprisingly simple result is 
that the derivatives of the eigenvalues are given by the expression 
\begin{equation}
\frac{d\Tk{i}}{dx} = \left|\Tilde{U}_{ei}\right|^{2}\,\frac{dV_{e}}{dx}
+\left|\Tilde{U}_{\mu i}\right|^{2}\,\frac{dV_{\mu}}{dx} + \left|\Tilde{U}_{\tau
i}\right|^{2}\,\frac{dV_{\tau }}{dx}.
\end{equation}
By unitarity then
$d\Tk{1}/dx+d\Tk{2}/dx+d\Tk{3}/dx=dV_{e}/dx+dV_{\mu}/dx+dV_{\tau}/dx$ which 
is also an identity that stems from the invariance of the trace of the
Hamiltonian. 
With this result we find  
\begin{widetext}
\begin{subequations}
\begin{eqnarray}
\frac{d\Tt{12}}{dx} & = & -\frac{1}{\dTk{12}\dTk{13}\dTk{23}}\,\left[
\Tc{12}\Ts{12}\dTk{13}\dTk{23}\left(
\left(\Tc{23}^{2}-\Ts{13}^{2}\Ts{23}^{2}\right)\,\frac{d\dV{e\mu}}{dx}
+\left(\Ts{23}^{2}-\Ts{13}^{2}\Tc{23}^{2}\right)\,\frac{d\dV{e\tau}}{dx} \right)
\right. \nonumber \\
&& \left. \qquad -\Tc{12}\Ts{12}\Ts{13}^{2}\dTk{12}^{2}\left(
\Ts{23}^{2}\frac{d\dV{e\mu}}{dx} +\Tc{23}^{2}\frac{d\dV{e\tau}}{dx} \right) -
\Ts{13}\Tc{23}\Ts{23}\Tce\,\left(\Tc{12}^{2}\dTk{13}^{2}-\Ts{12}^{2}\dTk{23}^{2}
\right)\,\frac{d\dV{\mu\tau}}{dx} \right], \label{eq:dt12dx}\\
\frac{d\Tt{13}}{dx} & = & -\frac{1}{\dTk{13}\dTk{23}}\,\left[
\Tc{13}\Ts{13}\left(
\Ts{12}^{2}\dTk{13}+\Tc{12}^{2}\dTk{23}\right)\left(\Ts{23}^{2}\frac{d\dV{e\mu}}
{dx}+\Tc{23}^{2}\frac{d\dV{e\tau}}{dx}\right)
+\Tc{12}\Ts{12}\Tc{13}\Tc{23}\Ts{23}\Tce\dTk{12}\frac{d\dV{\mu\tau}}{dx}
\right], \label{eq:dt13dx} \\
\frac{d\Tt{23}}{dx} & = &
-\Tc{12}\Ts{12}\Ts{13}\Tce\,\frac{\dTk{12}}{\dTk{13}\dTk{23}}\,\left(
\Ts{23}^{2}\frac{d\dV{e\mu}}{dx} +\Tc{23}^{2}\frac{d\dV{e\tau}}{dx} \right)
-
\frac{\Tc{23}\Ts{23}}{\dTk{13}\dTk{23}}\,\left(\Tc{12}^{2}\dTk{13}+\Ts{12}^{2}
\dTk{23} \right)\,\frac{d\dV{\mu\tau}}{dx}, \label{eq:dt23dx} 
\\
%%\end{eqnarray}\begin{eqnarray}
\frac{d\Tb}{dx} & = &
\frac{\Tc{12}\Ts{12}\Ts{13}\Tse}{\Tc{23}\Ts{23}}\,\frac{\dTk{12}}{\dTk{13}\dTk{
23}}\,\left( \Ts{23}^{2}\frac{d\dV{e\mu}}{dx} +\Tc{23}^{2}\frac{d\dV{e\tau}}{dx}
\right), \label{eq:dbetadx} 
\\
%%\end{eqnarray}\begin{eqnarray}
\frac{d\Td}{dx} & = &
-\frac{\Tc{12}\Ts{12}\Tc{23}\Tse}{\Ts{13}\Ts{23}}\,\frac{\dTk{12}}{\dTk{13}\dTk{
23}}\,\left[
\Ts{13}^{2}\,\left(\Ts{23}^{2}\frac{d\dV{e\mu}}{dx}+\Tc{23}^{2}\frac{d\dV{e\tau}
}{dx}\right) +\Ts{23}^{2}\frac{d\dV{\mu\tau}}{dx}\right], \label{eq:ddeltadx} \\
\frac{d\Te}{dx} & = &
-\frac{\Tc{12}\Ts{12}\Ts{13}\Tse}{\Tc{23}\Ts{23}}\,\frac{\dTk{12}}{\dTk{13}\dTk{
23}}\,\left(\Ts{23}^{2}\frac{d\dV{e\mu}}{dx}-\Tc{23}^{2}\frac{d\dV{e\tau}}{dx}
\right) \nonumber \\
&& \qquad +\frac{\Tc{23}\Ts{23}\Tse}{\Tc{12}\Ts{12}\Ts{13}}\,\frac{1}{\dTk{12}\dTk{13}\dTk{
23}}\,\left( \Tc{12}^{2}\Ts{12}^{2}\Tc{13}^{2}\dTk{12}^{2} -
\Tc{12}^{2}\Ts{13}^{2}\dTk{13}^{2}-\Ts{12}^{2}\Ts{13}^{2}\dTk{23}^{2}
\right)\,\frac{d\dV{\mu\tau}}{dx}. \label{eq:depsilondx} 
\end{eqnarray}
\end{subequations}
%%\end{widetext}
%%
Note that $d\Tb/dx$, $d\Td/dx$ and $d\Te/dx$ are all proportional to $\sin\Te$
and it is the appearance of this term in equation (\ref{eq:depsilondx}) that validates our
previous statement that $\Te=0$ for all $V$ if $\epsilon=0$. One also sees that when $\delta
V_{\mu\tau}=0$ there is a very simple relationship between $d\Tb/dx$, $d\Td/dx$,
$d\Te/dx$ and $d\Tt{23}/dx$ the last of which can be easily solved to give the 
Toshev identity \cite{T91}.

So we have finally reached the point where we are able to show that the matrix
$\TU^{\dag}\,d\TU/dx$ has the form
%%
%%\begin{widetext}
\begin{eqnarray}
\TU^{\dag}\,\frac{d\TU}{dx} & = &
\frac{d\Tt{12}}{dx}\,\left(\begin{array}{lll} 0 & e^{\idTa{12}} & 0 \\
-e^{-\idTa{12}} & 0 & 0 \\ 0 & 0 & 0 \end{array}\right) 
+ \frac{d\Tt{13}}{dx}\,\left(\begin{array}{lll} 0 & 0 & \Tc{12}\,e^{\idTa{13}}
\\ 0 & 0 & \Ts{12}\,e^{\idTa{23}} \\ -\Tc{12}\,e^{-\idTa{13}} &
-\Ts{12}\,e^{-\idTa{23}} & 0 \end{array}\right) 
\nonumber \\ &&  %%\qquad
+\frac{d\Tt{23}}{dx}\,
\left(\begin{array}{lll} -2\,\imath\,\Tc{12}\Ts{12}\Ts{13}\Tse &
\Ts{13}\,(\Tc{12}^{2}e^{\iTe}+\Ts{12}^{2}e^{-\iTe})\,e^{\idTa{12}} &
-\Ts{12}\Tc{13}\,e^{\idTa{13}-\iTe} \\
-\Ts{13}\,(\Tc{12}^{2}e^{-\iTe}+\Ts{12}^{2}e^{\iTe})\,e^{-\idTa{12}} &
2\,\imath\,\Tc{12}\Ts{12}\Ts{13}\Tse & \Tc{12}\Tc{13}\,e^{\idTa{23}-\iTe} \\ 
\Ts{12}\Tc{13}\,e^{-\idTa{13}+\iTe} & -\Tc{12}\Tc{13}\,e^{-\idTa{23}+\iTe} & 0
\end{array}\right) 
\nonumber \\ && %%\qquad
+\imath\,\frac{d\Tb}{dx}\,\Tilde{u}^{\dagger}_{\beta}\,\otimes\,\Tilde{u}_{\beta}
+\imath\,\frac{d\Td}{dx}\,\left(\begin{array}{lll} 1-\Tc{12}^{2}\Tc{13}^{2} &
-\Tc{12}\Ts{12}\Tc{13}^{2}e^{\idTa{12}} & -\Tc{12}\Tc{13}\Ts{13}e^{\idTa{13}} \\
-\Tc{12}\Ts{12}\Tc{13}^{2}e^{-\idTa{12}} & 1-\Ts{12}^{2}\Tc{13}^{2} &
-\Ts{12}\Tc{13}\Ts{13}e^{\idTa{23}} \\ -\Tc{12}\Tc{13}\Ts{13}e^{-\idTa{13}} &
-\Ts{12}\Tc{13}\Ts{13}e^{-\idTa{23}} & \Tc{13}^{2} \end{array}\right) 
\nonumber \\ && %%\qquad
+\imath\,\frac{d\Te}{dx}\,\left(\begin{array}{lll} \Ts{12}^{2} &
-\Tc{12}\Ts{12}e^{\idTa{12}} & 0 \\ -\Tc{12}\Ts{12}e^{-\idTa{12}} & \Tc{12}^{2}
& 0 \\ 0 & 0 & 0 \end{array}\right) 
%%\nonumber \\ && \qquad
-\imath\,\frac{d\Ta{1}}{dx}\left(\begin{array}{lll} 1 & 0 & 0 \\ 0 & 0 & 0
\\ 0 & 0 & 0 \end{array}\right)
    -\imath\,\frac{d\Ta{2}}{dx}\left(\begin{array}{lll} 0 & 0 & 0 \\ 0 & 1 & 0
\\ 0 & 0 & 0 \end{array}\right)
    -\imath\,\frac{d\Ta{3}}{dx}\left(\begin{array}{lll} 0 & 0 & 0 \\ 0 & 0 & 0
\\ 0 & 0 & 1 \end{array}\right)
\nonumber \\ &&
\end{eqnarray}
%%\end{widetext}
%%
where the vector $u_{\beta}$ is given by
%%
%%\begin{widetext}
\begin{equation}
u_{\beta}
=\left(\begin{array}{lll}\left(\Ts{12}\Ts{23}e^{\iTe}-\Tc{12}\Ts{13}\Tc{23}
\right)e^{-\iTa{1}}, &
-\left(\Tc{12}\Ts{23}e^{\iTe}+\Ts{12}\Ts{13}\Tc{23}\right)e^{-\iTa{2}}, &
\Tc{13}\Tc{23}e^{-\iTa{3}} \end{array}\right).
\end{equation}
\end{widetext}
So far $\Ta{1}$, $\Ta{2}$ and $\Ta{3}$ are unconstrained which permits us to
choose anything for them. 
It is at this point that we make our selection that $\Ta{1}$, $\Ta{2}$ and
$\Ta{3}$ be constants. 
 Since when using the scattering matrix approach, it is convenient to work with
a Hamilitonian
that has zeros along the diagonal,
another, sensible, alternative would be to assign $\Ta{1}$, $\Ta{2}$ and
$\Ta{3}$ so that the diagonal elements 
of $\TU^{\dag}\,d\TU/dx$ were exactly zero i.e. the contributions from
$d\Ta{1}/dx$, $d\Ta{2}/dx$ and 
$d\Ta{3}/dx$ in the equation above exactly canceled the contributions from
$d\Tt{23}/dx$, $d\Tb/dx$, $d\Td/dx$ and 
$d\Te/dx$ that appear along the diagonal. But, as we shall 
shortly show in sec \ref{sec:adiabatic}, our desire 
to cancel the diagonal elements of $\TU^{\dag}\,d\TU/dx$ can be achieved 
more conveniently by another route hence we opt to set the three phases to be
constants. \\ \\

%%%%%%%%%%%%%%%%%%%%%%%%%%%%%%%%%%%%%%%%%%%%%%%%%%%%%%%%%%

\subsection{The High Density Limit} \label{sec:limit}

In the limit of high density, the expressions for the matter angles and
phases, (\ref{eq:theta12}) to (\ref{eq:delta2}), become difficult to work with. 
The on-diagonal components of the flavor space Hamiltonian, 
$\TH_{ee}$, $\TH_{\mu \mu}$ and $\TH_{\tau \tau}$, are dominated by the matter
potentials which are much larger than the off-diagonal components. Further, the eigenvalues
$\Tk{1}$, $\Tk{2}$, and $\Tk{3}$ are also dominated by the matter potentials so when using the expressions 
(\ref{eq:theta12}) to (\ref{eq:delta2}) one is sometimes in the situation where
two large numbers must be subtracted to find a small difference. For this reason it is
useful to find expansions for the eigenvalues in powers of the potentials.
In this section we derive the asymptotic limits of the neutrino and 
antineutrino eigenvalues when $V_{e}\rightarrow \infty$ assuming 
$V_{e} \gg V_{\mu}$ and $V_{e} \gg V_{\tau}$.

We begin by considering the two quantities $\TQ$ and $\TR$. The original expressions
for these quantities may be rewritten as 
\begin{widetext}
\begin{eqnarray}
\TQ & = & \frac{-V^{2}_{e}}{9}\,\left[1
-\frac{(18\,q_{e}+V_{\mu}+V_{\tau})}{V_{e}}\,+\frac{\left(V_{\mu}^{2}-V_{\mu}V_{
\tau}+V_{\tau}^{2}-9\,\left[Q+2\,q_{\mu}V_{\mu}+2\,q_{\tau}V_{\tau}\right]
\right)}{V_{e}^{2}}\right],\\
& \equiv & \frac{-V^{2}_{e}}{9}\,\left[
1+\frac{a_{Q}}{V_{e}}+\frac{b_{Q}}{V_{e}^{2}} \right],\\
\TR & = & \frac{V^{3}_{e}}{27}\,\left[1
-\frac{3\,(18\,q_{e}+V_{\mu}+V_{\tau})}{2\,V_{e}}\,
-\frac{3\,\left(V_{\mu}^{2}-4V_{\mu}V_{\tau}+V_{\tau}^{2}-18\,\left[r_{e}-2\,q_{
\mu}V_{\tau}-2\,q_{\tau}V_{\mu}\right]\right)}{2\,V_{e}^{2}} \right.\nonumber \\
&& \left.
+\frac{\left(2V_{\mu}^{3}-3V_{\mu}^{2}V_{\tau}-3V_{\mu}V_{\tau}^{2}+2V_{\tau}^{3
}+54\,\left[R+r_{\mu}V_{\mu}+r_{\tau}V_{\tau}-2q_{e}V_{\mu}V_{\tau}-q_{\mu}
V_{\mu}^{2}-q_{\tau}V_{\tau}^{2}\right]\right)}{2\,V_{e}^{3}}\right], \\
& \equiv & \frac{V^{3}_{e}}{27}\,\left[
1+\frac{3\,a_{Q}}{2\,V_{e}}+\frac{b_{R}}{V_{e}^{2}}+\frac{c_{R}}{V_{e}^{3}}
\right].
\end{eqnarray}
%%\end{widetext}
%%
For antineutrinos
%%
%%\begin{widetext}
\begin{eqnarray}
\TQbar & = & \frac{-V^{2}_{e}}{9}\,\left[1
+\frac{(18\,q_{e}-V_{\mu}-V_{\tau})}{V_{e}}\,+\frac{\left(V_{\mu}^{2}-V_{\mu}V_{
\tau}+V_{\tau}^{2}-9\,\left[Q-2\,q_{\mu}V_{\mu}-2\,q_{\tau}V_{\tau}\right]
\right)}{V_{e}^{2}}\right],\\
& \equiv & \frac{-V^{2}_{e}}{9}\,\left[
1+\frac{\bar{a}_{Q}}{V_{e}}+\frac{\bar{b}_{Q}}{V_{e}^{2}} \right],\\
\TRbar & = & -\frac{V^{3}_{e}}{27}\,\left[1
+\frac{3\,(18\,q_{e}-V_{\mu}-V_{\tau})}{2\,V_{e}}\,
-\frac{3\,\left(V_{\mu}^{2}-4V_{\mu}V_{\tau}+V_{\tau}^{2}-18\,\left[r_{e}+2\,q_{
\mu}V_{\tau}+2\,q_{\tau}V_{\mu}\right]\right)}{2\,V_{e}^{2}} \right.\nonumber \\
&& \left.
+\frac{\left(2V_{\mu}^{3}-3V_{\mu}^{2}V_{\tau}-3V_{\mu}V_{\tau}^{2}+2V_{\tau}^{3
}-54\,\left[R-r_{\mu}V_{\mu}-r_{\tau}V_{\tau}-2q_{e}V_{\mu}V_{\tau}-q_{\mu}
V_{\mu}^{2}-q_{\tau}V_{\tau}^{2}\right]\right)}{2\,V_{e}^{3}}\right], \\
& \equiv & -\frac{V^{3}_{e}}{27}\,\left[
1+\frac{3\,\bar{a}_{Q}}{2\,V_{e}}+\frac{\bar{b}_{R}}{V_{e}^{2}}+\frac{\bar{c}_{R
}}{V_{e}^{3}} \right].
\end{eqnarray}
\end{widetext}
Using these expressions we find that 
$\cos\omega \rightarrow 1$, i.e. $\omega\rightarrow 0$ in the $V_{e}\rightarrow
\infty$ limit. If we expand out $\cos\omega$ in terms of powers of $\omega$ then
\begin{equation}
1 -\frac{\omega^{2}}{2}+\frac{\omega^{4}}{24} + \dots= \frac{
1+a_{R}/V_{e}+b_{R}/V_{e}^{2}+c_{R}/V_{e}^{3} }{\left(
1+a_{Q}/V_{e}+b_{Q}/V_{e}^{2}\right)^{3/2}}.
\end{equation}
Solving for $\omega$ and expanding out the right-hand side of the resulting 
expression up to order $1/V_{e}^{2}$ we are eventually led to 
\begin{subequations}
\begin{eqnarray}
\omega & =& \frac{\sqrt{3\,a_{Q}^{2}+12\,b_{Q}-8\,b_{R}}}{2\,V_{e}}
%%\nonumber \\ && \qquad
-\frac{\left[5\,a_{Q}^{3}+12\,a_{Q}\,(b_{Q}-b_{R})+8\,c_{R}\right]}{4\,V^{2}_{e}
\,\sqrt{3\,a_{Q}^{2}+12\,b_{Q}-8\,b_{R}}}, 
\\
%%\end{eqnarray}\begin{eqnarray}
& \equiv & \frac{a_{\omega}}{V_{e}}+\frac{b_{\omega}}{V_{e}^{2}}
\end{eqnarray}
\end{subequations}
which defines the quantities $a_{\omega}$ and $b_{\omega}$.
When we substitute in the definitions of $a_{Q}$, $b_{Q}$ etc. we find that 
%%
%%\begin{widetext}
\begin{subequations}
\begin{eqnarray}
a_{\omega} & = &
\frac{3\,\sqrt{3}}{2}\sqrt{4\,|\Hmt|^{2}+\left(\THmm-\THtt\right)^{2}} 
\\
%%\end{eqnarray}\begin{widetext}\begin{eqnarray}
b_{\omega} & = &
\frac{3\,\sqrt{3}}{4}\,\left[\THmm+\THtt-2\,H_{ee}\right]\,\sqrt{4\,|\Hmt|^{2}
+\left(\THmm-\THtt\right)^{2}} 
\nonumber \\ && \qquad
-\frac{3\,\sqrt{3}}{2}\frac{\left[
2\left(\Hem\Hmt\Hte+\Het\Htm\Hme\right)+\left(|\Hem|^{2}-|\Het|^{2}
\right)\left(\THmm-\THtt\right)\right]}{\sqrt{4|\Hmt|^{2}
+\left(\THmm-\THtt\right)^{2}}}. 
\end{eqnarray}
%%\end{widetext}
\end{subequations}

For antineutrinos we find instead that $\cos\bar{\omega}\rightarrow -1$ which
indicates the angle $\bar{\omega}$ approaches $\pi$ as $V_{e} \rightarrow \infty$. In this
case we instead expand $\cos\bar{\omega}$ 
in terms of $\bar{\omega}-\pi$ and find the asymptotic limit is
\begin{equation}
\bar{\omega} = \pi-
\frac{\bar{a}_{\omega}}{V_{e}}-\frac{\bar{b}_{\omega}}{V_{e}^{2}}.
\end{equation}
where $\bar{a}_{\omega}$ and $\bar{b}_{\omega}$ are given by exactly the same 
expressions as $a_{\omega}$ and $b_{\omega}$, although the elements of the
Hamiltonian are, in this case, taken from the Hamiltonian appropriate for antineutrinos. 
With the asymptotic expression for $\omega$ and $\bar{\omega}$ determined we can
then proceed to the eigenvalues.
For neutrinos these are given by
\begin{eqnarray}
\Tk{i}&=&\frac{\left(T+V_{e}+V_{\mu}+V_{\tau}\right)}{3}
%%\nonumber \\ && 
+\frac{2\,V_{e}}{3}\,\cos\left(\frac{\omega+\omega_{i}}{3}\right)\,\sqrt{1+a_{Q}
/V_{e}+b_{Q}/V_{e}^{2}}.
%%\nonumber \\ && 
\end{eqnarray}
Using the expression for the expansion of $\omega$ then, up to $1/V_{e}$, we
eventually find
\begin{widetext}
\begin{subequations}
\begin{eqnarray}
\Tk{i} &
=&\frac{4\,V_{e}}{3}\,\cos\left(\frac{\omega_{i}-\pi}{6}\right)\,\cos\left(\frac
{\omega_{i}+\pi}{6}\right) 
+\frac{T+V_{\mu}+V_{\tau}}{3} 
%%\nonumber \\ &&
+\frac{a_{Q}}{3}\,\cos\left(\frac{\omega_{i}}{3}\right)
-\frac{2\,a_{\omega}}{9}\,\sin\left(\frac{\omega_{i}}{3}\right) 
\nonumber \\ && \qquad
- \frac{1}{9\,V_{e}}\left[
\left(\frac{3\,a_{Q}^{2}}{4}-3\,b_{Q}+\frac{a_{\omega}^{2}}{3}
\right)\cos\left(\frac{\omega_{i}}{3}\right)
%%\right. \nonumber \\ && \left. \qquad
+\left(a_{Q}a_{\omega}+2\,b_{\omega}\right)\sin\left(\frac{\omega_{i}}{3}\right)
\right], \\
& \equiv &
\frac{4\,V_{e}}{3}\,\cos\left(\frac{\omega_{i}-\pi}{6}\right)\,\cos\left(\frac{
\omega_{i}+\pi}{6}\right) + C_{i} +\frac{a_{i}}{V_{e}}.
\end{eqnarray}
\end{subequations}
%%\end{widetext}
%%
Notice that if $\omega_{i} = 2\pi$ or $\omega_{i} = 4\pi$ 
then the term linear in $V_{e}$ vanishes: it only survives when $\omega_{i}=0$. 
Substituting in for $a_{\omega}$ etc. we find 
%%
%%\begin{widetext}
\begin{subequations}
\begin{eqnarray}
C_{i}& = & \frac{H_{ee}}{3}\left[
1+2\,\cos\left(\frac{\omega_{i}}{3}\right)\right] 
+\left(\frac{\THmm+\THtt}{3}\right)\left[1-\cos\left(\frac{\omega_{i}}{3}
\right)\right] 
-\frac{\sqrt{4\,|\Hmt|^{2}+(\THmm-\THtt)^{2}}}{\sqrt{3}}\,\sin\left(\frac{
\omega_{i}}{3}\right) ,\\
a_{i} & = &
\frac{\left[2\left(\Hem\Hmt\Hte+\Het\Htm\Hme\right)+\left(|\Hem|^{2}
-|\Het|^{2}\right)\left(\THmm-\THtt\right)\right]}{\sqrt{3}\,\sqrt{4|\Hmt|^{2}
+\left(\THmm-\THtt\right)^{2}}}\,\sin\left(\frac{\omega_{i}}{3}\right)
\nonumber \\&& \qquad
+\left(|\Hem|^{2}+|\Het|^{2}\right)\,\cos\left(\frac{\omega_{i}}{3}\right)
\end{eqnarray}
\end{subequations}
%%\end{widetext}
%%
Note that when $\omega_{i}=0$ then $C_{i}=H_{ee}$ so that the first two terms of
this eigenvalue $k_{i}$ are $k_{i}=V_{e}+H_{ee} = \THee$.

The expression for the eigenvalues in the case of antineutrinos is instead 
\begin{eqnarray}
\Tk{i}&=&\frac{\left(T-V_{e}-V_{\mu}-V_{\tau}\right)}{3}
%%\nonumber \\ && 
+\frac{2\,V_{e}}{3}\,\cos\left(\frac{\bar{\omega}+\omega_{i}}{3}\right)\,\sqrt{
1+\bar{a}_{Q}/V_{e}+\bar{b}_{Q}/V_{e}^{2}}.
%%\nonumber \\ && 
\end{eqnarray}
When we substitute in the asymptotic expansion for $\bar{\omega}$ we find
%%
%%\begin{widetext}
\begin{subequations}
\begin{eqnarray}
\Tk{i} 
&=&-\frac{4\,V_{e}}{3}\,\cos\left(\frac{\bar{\omega}_{i}-\pi}{6}\right)\,
\cos\left(\frac{\bar{\omega}_{i}+3\,\pi}{6}\right) 
+\frac{(T-V_{\mu}-V_{\tau})}{3} 
%%\nonumber \\ && 
+\frac{\bar{a}_{Q}}{3}\,\cos\left(\frac{\omega_{i}+\pi}{3}\right)
+\frac{2\,\bar{a}_{\omega}}{9}\,\sin\left(\frac{\omega_{i}+\pi}{3}\right)
\nonumber \\ && \qquad
-\frac{1}{9\,V_{e}}\left[\left(\frac{3\,\bar{a}_{Q}^{2}}{4}-3\,\bar{b}_{Q}+\frac
{\bar{a}_{\omega}^{2}}{3}\right)\cos\left(\frac{\omega_{i}+\pi}{3}\right)
%%\right. \nonumber \\ && \qquad \left.
-\left(\bar{a}_{Q}\bar{a}_{\omega}+2\,\bar{b}_{\omega}\right)\sin\left(\frac{
\omega_{i}+\pi}{3}\right) \right] ,\\
& \equiv &
-\frac{4\,V_{e}}{3}\,\cos\left(\frac{\bar{\omega}_{i}-\pi}{6}\right)\,
\cos\left(\frac{\bar{\omega}_{i}+3\,\pi}{6}\right) + \bar{C}_{i}
+\frac{\bar{a}_{i}}{V_{e}}.
\end{eqnarray}
\end{subequations}
%%\end{widetext}
%%
Here the linear term in $V_{e}$ vanishes if $\omega_{i}\neq 2\pi$. 
In this case the expression $\bar{C}_{i}$ is equal to 
%%
%%\begin{widetext}
\begin{subequations}
\begin{eqnarray}
\bar{C}_{i}
&=&\frac{\THbaree}{3}\left[
1-2\,\cos\left(\frac{\bar{\omega}_{i}+\pi}{3}\right)\right] 
%%\nonumber \\ &&
+\left(\frac{\THbarmm+\THbartt}{3}\right)\left[
1+\cos\left(\frac{\bar{\omega}_{i}+\pi}{3}\right)\right] 
\nonumber \\ && \qquad
+ \frac{\sqrt{4\,|\Hbarmt|^{2}+(\THbarmm-\THbartt)^{2}} }{\sqrt{3}}\,\sin\left(\frac{\bar{\omega}_{i}+\pi}{3}\right),
\\
\bar{a}_{i}& = &
\frac{\left[2\left(\Hbarem\Hbarmt\Hbarte+\Hbaret\Hbartm\Hbarme\right)+\left(|\Hbarem|^{2}-|\Hbaret|^{2}\right) \left(\THbarmm-\THbartt\right)\right]}{\sqrt{3}\,\sqrt{4|\Hbarmt|^{2}+\left(\THbarmm-\THbartt\right)^{2}}}\,
\sin\left(\frac{\omega_{i}+\pi}{3}\right) 
\nonumber \\ && \qquad
+\left(|\Hbarem|^{2}+|\Hbaret|^{2}\right)\cos\left(\frac{
\omega_{i}+\pi}{3}\right).
\end{eqnarray}
\end{subequations}
\end{widetext}

We can now write down the eigenvalues in the high density limit.
For neutrinos and the normal hierarchy $\omega_{1}=2\pi$, $\omega_{2}=4\pi$ and
$\omega_{3}=0$ so 
in the infinite density limit 
\begin{subequations}
\begin{eqnarray}
\Tk{1} &\rightarrow&
\frac{\THmm+\THtt}{2}-\sqrt{|\Hmt|^{2}+\frac{\left(\THmm-\THtt\right)^{2}}{4}}
%%\nonumber \\ && \qquad 
+\frac{a_{1}}{V_{e}} + {\cal O}(1/V_{e}^2) ,\\
\Tk{2} &\rightarrow&
\frac{\THmm+\THtt}{2}+\sqrt{|\Hmt|^{2}+\frac{\left(\THmm-\THtt\right)^{2}}{4}}
%%\nonumber \\ && \qquad 
+\frac{a_{2}}{V_{e}} + {\cal O}(1/V_{e}^2) ,\\
\Tk{3} &\rightarrow& \THee+\frac{a_{3}}{V_{e}} + {\cal O}(1/V_{e}^2).
\end{eqnarray}
\end{subequations}
For antineutrinos in the normal hierarchy
\begin{subequations}
\begin{eqnarray}
\Tk{1} &\rightarrow& \THbaree+\frac{\bar{a}_{1}}{V_{e}} + {\cal
O}(1/V_{e}^2),\\
\Tk{2} &\rightarrow&
\frac{\THbarmm+\THbartt}{2}-\sqrt{|\Hbarmt|^{2
}+\frac{\left(\THbarmm-\THbartt\right)^{2}}{4}}
%%\nonumber \\ && \qquad 
+\frac{\bar{a}_{2}}{V_{e}} + {\cal O}(1/V_{e}^2),\\
\Tk{3} &\rightarrow&
\frac{\THbarmm+\THbartt}{2}+\sqrt{|\Hbarmt|^{2
}+\frac{\left(\THbarmm-\THbartt\right)^{2}}{4}}
%%\nonumber \\ && \qquad 
+\frac{\bar{a}_{3}}{V_{e}} + {\cal O}(1/V_{e}^2).
\end{eqnarray}
\end{subequations}
For neutrinos and an inverted hierarchy $\omega_{1}=4\pi$, $\omega_{2}=0$ and
$\omega_{3}=2\pi$ 
\begin{subequations}
\begin{eqnarray}
\Tk{1} &\rightarrow&
\frac{\THmm+\THtt}{2}+\sqrt{|\Hmt|^{2}+\frac{\left(\THmm-\THtt\right)^{2}}{4}}
%%\nonumber \\ && \qquad 
+\frac{a_{1}}{V_{e}} + {\cal O}(1/V_{e}^2),\\
\Tk{2} &\rightarrow& \THee+\frac{a_{2}}{V_{e}} + {\cal O}(1/V_{e}^2),\\
\Tk{3} &\rightarrow&
\frac{\THmm+\THtt}{2}-\sqrt{|\Hmt|^{2}+\frac{\left(\THmm-\THtt\right)^{2}}{4}}
%%\nonumber \\ && \qquad 
+\frac{a_{3}}{V_{e}} + {\cal O}(1/V_{e}^2).
\end{eqnarray}
\end{subequations}
And for antineutrinos in the inverted hierarchy
\begin{subequations}
\begin{eqnarray}
\Tk{1} &\rightarrow&
\frac{\THbarmm+\THbartt}{2}-\sqrt{|\Hbarmt|^{2
}+\frac{\left(\THbarmm-\THbartt\right)^{2}}{4}}
%%\nonumber \\ && \qquad 
+\frac{\bar{a}_{1}}{V_{e}} + {\cal O}(1/V_{e}^2),\\
\Tk{2} &\rightarrow&
\frac{\THbarmm+\THbartt}{2}+\sqrt{|\Hbarmt|^{2
}+\frac{\left(\THbarmm-\THbartt\right)^{2}}{4}}
%%\nonumber \\ && \qquad 
+\frac{\bar{a}_{2}}{V_{e}} + {\cal O}(1/V_{e}^2),\\
\Tk{3} &\rightarrow& \THbaree+\frac{\bar{a}_{3}}{V_{e}} + {\cal O}(1/V_{e}^2).
\end{eqnarray}
\end{subequations}
The expressions we have derived for $\Tk{1}$, $\Tk{2}$ and $\Tk{3}$ in the
asymptotic limit can be employed together with equations (\ref{eq:theta12}) to (\ref{eq:delta2})
to more easily determine the correct values for the mixing angles and phases in situations where the matter potential
is large.

%%%%%%%%%%%%%%%%%%%%%%%%%%%%%%%%%%%%%%%%%%%%%%%%%%%%%%%%%%%%%%%%%%%%%%%%
%%%%%%%%%%%%%%%%%%%%%%%%%%%%%%%%%%%%%%%%%%%%%%%%%%%%%%%%%%%%%%%%%%%%%%%%
%%%%%%%%%%%%%%%%%%%%%%%%%%%%%%%%%%%%%%%%%%%%%%%%%%%%%%%%%%%%%%%%%%%%%%%%

\section{The Adiabatic Basis}\label{sec:adiabatic}

Returning to the problem of neutrino propagation we 
now introduce a new basis $\psi^{(a)}$, which we call the adiabatic basis, with
$\psi^{(m)} = W(x)\psi^{(a)}$.

After making this change of basis we find the Schrodinger equation has become 
\begin{subequations}
\begin{eqnarray}
&&\imath \frac{d\psi^{(a)}}{dx} 
%%\nonumber \\ && \qquad
= \left( W^{\dag}\Tilde{K}W - \imath W^{\dag}\frac{dW}{dx} -\imath W^{\dag}\TU^{\dag}\frac{d\TU}{dx}W
\right)\,\psi^{(a)}, 
\nonumber \\ && 
\\ && \qquad \equiv  \TH^{(a)}\psi^{(a)}.
\end{eqnarray}
\end{subequations}
and we choose $W$ so that it removes the diagonal of $\TH^{(a)}$.
The matrix $W$ is simply
\begin{equation}
W=\left(\begin{array}{lll} \exp[-2\,\imath \pi \phi_{1}] & 0 & 0 \\ 0 &
\exp[-2\,\imath \pi \phi_{2}] & 0 \\ 0 & 0 & \exp[-2\,\imath \pi \phi_{3}]
\end{array}\right) 
\label{eq:W}
\end{equation}
where $\phi_{1}$, $\phi_{2}$ and $\phi_{3}$ are defined to be
\begin{subequations}
\begin{eqnarray}
\frac{d\phi_{1}}{dx} & =
&\frac{1}{2\,\pi}\left(\Tk{1}-\frac{\Ts{12}\Ts{13}\Tc{23}\Ts{23}\Tse}{\Tc{12}}
\frac{\dTk{23}}{\dTk{12}\,\dTk{13}}\frac{d\dV{\mu\tau}}{dx} \right),
%%\nonumber \\&&
\label{eq:dphi12dx}
\\
%%\end{eqnarray}\begin{eqnarray}
\frac{d\phi_{2}}{dx} & =
&\frac{1}{2\,\pi}\left(\Tk{2}-\frac{\Tc{12}\Ts{13}\Tc{23}\Ts{23}\Tse}{\Ts{12}}
\,\frac{\dTk{13}}{\dTk{12}\,\dTk{23}}\frac{d\dV{\mu\tau}}{dx} \right),
%%\nonumber \\ && 
\label{eq:dphi13dx}\\
\frac{d\phi_{3}}{dx} & = 
&\frac{1}{2\,\pi}\left(\Tk{3}-\frac{\Tc{12}\Ts{12}\Tc{13}^{2}\Tc{23}\Ts{23}
\Tse}{\Ts{13}}\frac{\dTk{12}}{\dTk{13}\,\dTk{23}}\frac{d\dV{\mu\tau}}{dx}
\right).
%%\nonumber \\&&
\label{eq:dphi23dx}
\end{eqnarray}
\end{subequations}
If we had used $\Ta{1}$, $\Ta{2}$ and $\Ta{3}$ to cancel off the diagonal
elements of $\TU^{\dag}\,d\TU/dx$ then 
equations (\ref{eq:dphi12dx}) to (\ref{eq:dphi23dx}) would  only contain the
first terms in the above expression, i.e. only the contribution proportional to
$\Tk{1}$, $\Tk{2}$ and $\Tk{3}$. The name ``adiabatic'' comes from the
consideration of the two-flavor oscillation problem where the phase analogous to
the $\phi_{i}$'s is the adiabatic phase of the matter states. For three flavors 
since their are multiple phases which also contain additional terms
proportional to $d\dV{\mu\tau}/dx$ the situation is more complex. However,
this basis is  useful for providing insight into the phase effects, e.g.
 Ref \cite{K&M2006}, and the effects of the CP phase $\epsilon$  as
described in Sec. \ref{sec:identities}.  Further, it is an ideal basis in which to perform
neutrino flavor transformation calculations using the $S$ matrix prescription.
After removing the diagonal elements of $\TH^{(a)}$ we can write out $\TH^{(a)}$ as 
\begin{widetext}
\begin{equation}
\TH^{(a)} =
\left(\begin{array}{lll} 
0 & \imath\,\frac{\dTk{12}}{2\,\pi}\,\Gamma_{12}\,e^{2\,\imath\pi\dphi{12}} &
\imath\,\frac{\dTk{13}}{2\,\pi}\,\Gamma_{13}\,e^{2\,\imath\pi\dphi{13}} \\ 
-\imath\,\frac{\dTk{12}}{2\,\pi}\,\Gamma^{\ast}_{12}\,e^{-2\,\imath\pi\dphi{12}}
& 0 & \imath\,\frac{\dTk{23}}{2\,\pi}\,\Gamma_{23}\,e^{2\,\imath\pi\dphi{23}}
\\ 
-\imath\,\frac{\dTk{13}}{2\,\pi}\,\Gamma^{\ast}_{13}\,e^{-2\,\imath\pi\dphi{13}}
&
-\imath\,\frac{\dTk{23}}{2\,\pi}\,\Gamma^{\ast}_{23}\,e^{-2\,\imath\pi\dphi{23}}
& 0 
\end{array}\right) \label{eq:Ha}.
\end{equation}
%%\end{widetext}
%%
where, as usual, $\dphi{ij} = \phi_{i}-\phi_{j}$. This equation 
defines three functions $\Gamma_{12}, \Gamma_{13}$ and $\Gamma_{23}$ 
which are the non-adiabaticity parameters for 3 flavor neutrino oscillations. 
By matching the expressions we find 
%%
%%\begin{widetext}
\begin{subequations}
\begin{eqnarray}
\Gamma_{12} & = &
-\frac{2\,\pi\,e^{\idTa{12}}}{\dTk{12}}\,\left(\frac{d\Tt{12}}{dx}
+\Ts{13}\left(\Tc{12}^{2}e^{\iTe}+\Ts{12}^{2}e^{-\iTe}\right)\,\frac{d\Tt{23}}{
dx} \right. \nonumber \\
&& \left.
-\imath\left(\Ts{12}\Ts{23}e^{-\iTe}-\Tc{12}\Ts{13}\Tc{23}\right)\left(\Tc{12}
\Ts{23}e^{\iTe}+\Ts{12}\Ts{13}\Tc{23}\right)\,\frac{d\Tb}{dx}
-\imath\Tc{12}\Ts{12}\Tc{13}^{2}\,\frac{d\Td}{dx}
-\imath\Tc{12}\Ts{12}\,\frac{d\Te}{dx} \right), \\
\Gamma_{13} & = &
-\frac{2\,\pi\,e^{\idTa{13}}}{\dTk{13}}\,\left(\Tc{12}\,\frac{d\Tt{13}}{dx}
-\Ts{12}\Tc{13}e^{-\iTe}\,\frac{d\Tt{23}}{dx}
+\imath\left(\Ts{12}\Ts{23}e^{-\iTe}-\Tc{12}\Ts{13}\Tc{23}\right)\Tc{13}\Tc{23}\
,\frac{d\Tb}{dx} -\imath\Tc{12}\Tc{13}\Ts{13}\,\frac{d\Td}{dx} \right), \\
\Gamma_{23} & = &
-\frac{2\,\pi\,e^{\idTa{23}}}{\dTk{23}}\,\left(\Ts{12}\,\frac{d\Tt{13}}{dx}
+\Tc{12}\Tc{13}e^{-\iTe}\,\frac{d\Tt{23}}{dx}
-\imath\left(\Tc{12}\Ts{23}e^{-\iTe}+\Ts{12}\Ts{13}\Tc{23}\right)\Tc{13}\Tc{23}\
,\frac{d\Tb}{dx} -\imath\Ts{12}\Tc{13}\Ts{13}\,\frac{d\Td}{dx} \right).
\end{eqnarray}
\end{subequations}
%%\end{widetext}
%%
Note how the derivatives of the two phases $\Tb$ and $\Td$ appear in these
expressions. 
When we substitute in the expressions for the derivatives we find
%%
%%\begin{widetext}
\begin{subequations}
\begin{eqnarray}
\Gamma_{12} & = & \frac{2\,\pi\,e^{\idTa{12}}}{\dTk{12}^{2}}\,\left[
\Tc{12}\Ts{12}\Tc{13}^{2}\left( \Ts{23}^{2}\frac{d\dV{e\mu}}{dx}
+\Tc{23}^{2}\frac{d\dV{e\tau}}{dx} \right) 
-\left(
\Tc{12}\Ts{12}\left(\Tc{23}^{2}-\Ts{23}^{2}\right)+\Ts{13}\Tc{23}\Ts{23}
\left(\Tc{12}^{2}e^{\iTe}-\Ts{12}^{2}e^{-\iTe}\right)\right)\frac{d\dV{\mu\tau}}
{dx} \right], \nonumber \\
&& \label{eq:Gamma12}\\
\Gamma_{13} & = & \frac{2\,\pi\,e^{\idTa{13}}}{\dTk{13}^{2}}\left[
\Tc{12}\Tc{13}\Ts{13}\left( \Ts{23}^{2}\frac{d\dV{e\mu}}{dx}
+\Tc{23}^{2}\frac{d\dV{e\tau}}{dx} \right) 
-\Ts{12}\Tc{13}\Tc{23}\Ts{23}e^{-\iTe}\frac{d\dV{\mu\tau}}{dx}
\right],\label{eq:Gamma13}\\
\Gamma_{23} & = & \frac{2\,\pi\,e^{\idTa{23}}}{\dTk{23}^{2}}\left[
\Ts{12}\Tc{13}\Ts{13}\left( \Ts{23}^{2}\frac{d\dV{e\mu}}{dx}
+\Tc{23}^{2}\frac{d\dV{e\tau}}{dx} \right)
+\Tc{12}\Tc{13}\Tc{23}\Ts{23}e^{-\iTe}\frac{d\dV{\mu\tau}}{dx}
\right]\label{eq:Gamma23}.
\end{eqnarray}
\end{subequations}
\end{widetext}
If we restict ourselves to $\dV{\mu\tau}=0$ then we see that 
$\Gamma_{12} \propto \Tc{12}\Ts{12}/\dTk{12}^{2}\,dV_{e}/dx$ which is exactly
the two flavor non-adiabaticity parameter as described in 
\cite{K&M2006}. The other similar terms, 
$\Gamma_{13} \propto \Tc{13}\Ts{13}/\dTk{13}^{2}\,dV_{e}/dx$ and 
$\Gamma_{23} \propto \Tc{13}\Ts{13}/\dTk{23}^{2}\,dV_{e}/dx$ also take on
similar meanings. This Hamiltonian is constructed in such a way that,
similar to the two flavor case, the focus is on the ``non-adiabatic'' pieces of
the solution, i.e. the places near the resonances where the matter eigenstates are likely to 
swap. There are some corrections for the case of three flavors that come from the 
terms $\Tc{13}^{2}$, $\Tc{12}$ and $\Ts{12}$ and the three, arbitrary, complex exponentials in $\Gamma_{12},
\Gamma_{13}$ and $\Gamma_{23}$ respectively. It can be seen that it is the behavior of $\Tt{12}$ that selects between mixing
of states $\Tnu{1}$ and $\Tnu{3}$ or between $\Tnu{2}$ and $\Tnu{3}$.

%%%%%%%%%%%%%%%%%%%%%%%%%%%%%%%%%%%%%%%%%%%%%%%%%%%%%%%%%%%%%%%%%%%%%%%%
%%%%%%%%%%%%%%%%%%%%%%%%%%%%%%%%%%%%%%%%%%%%%%%%%%%%%%%%%%%%%%%%%%%%%%%%
%%%%%%%%%%%%%%%%%%%%%%%%%%%%%%%%%%%%%%%%%%%%%%%%%%%%%%%%%%%%%%%%%%%%%%%%

\section{The Scattering Matrix} \label{sec:to S}

The Schrodinger equation for the evolution of the neutrino wavefunction is
\begin{equation} 
\frac{d\psi}{dx} = -\imath\,\TH(x)\,\psi(x). \label{eq:SE}
\end{equation}
When we integrate equation (\ref{eq:SE}) we obtain 
\begin{equation} 
\psi(X) = \psi(X_{0}) -\imath\;\int_{X_{0}}^{X} \; \dx{1} \TH(\x{1})\;\psi(\x{1}).
\end{equation}
We choose the initial point to be at $X_{0}$. 
Repeated substitution of this result into itself yields 
\begin{widetext}
\begin{subequations}
\begin{eqnarray} 
\psi(X)
& = & \psi(X_{0}) -\imath\;\int_{X_{0}}^{X} \; \dx{1} \TH_{1} \,\psi(X_{0})
+  (-\imath)^{2}\;\int_{X_{0}}^{X} \;\dx{1} \; \TH_{1} \, \int_{X_{0}}^{\x{1}}
\; \dx{2} \;\TH_{2} \,\psi(X_{0}) +  \ldots ,\\
& = & \left\{ 1 -\imath\;\int_{X_{0}}^{X} \; \dx{1} \;\TH_{1} \, 
+  (-\imath)^{2}\; \int_{X_{0}}^{X} \; \dx{1} \; \TH_{1} \, \int_{X_{0}}^{\x{1}}
\; \dx{2} \;\TH_{2} \,
+  \ldots \right\} \; \psi(X_{0}) \label{eq:integral}
\end{eqnarray}
\end{subequations}
\end{widetext}
where the subscripts on the $\TH$'s mean $\TH_{i} = \TH(x_{i})$. 
This equation defines the scattering matrix $S(X,X_{0})$ since 
\begin{equation} 
\psi(X) =  S(X,X_{0}) \psi(X_{0}). \label{eq:psiX}
\end{equation}
The upper limits on the integrals appearing in equation (\ref{eq:integral})
indicate the 
space ordering but we can change all the upper limits to $X$ by using identities
such as 
\begin{eqnarray}
&& \int_{X_{0}}^{X} \dx{1} \; \TH_{1} \int_{X_{0}}^{\x{1}} \; \dx{2} \;\TH_{2} 
%%\nonumber \\ && \qquad 
= \frac{1}{2!} \,\int_{X_{0}}^{X}\;\dx{1}\,\int_{X_{0}}^{X}\;\dx{2}\, \left\{
\TH_{1}\,\TH_{2} \Theta(\x{1}-\x{2}) 
%%\right. \nonumber \\ && \qquad \qquad \left.
+ \TH_{2}\,\TH_{1} \Theta(\x{2}-\x{1}) \right\}
\end{eqnarray}
where $\Theta(\x{1}-\x{2})$ is the Heaviside step function. Using this result
and similar identities for
the higher order multiple integrals, allows us to write $S$ as 
\begin{eqnarray}
&&S(X,X_{0}) 
%%\nonumber \\ && \qquad
= 1 + (-\imath)\int_{X_{0}}^{X}\dx{1}\,\TH_{1} 
%%\nonumber \\ && \qquad \;
+ \frac{(-\imath)^{2}}{2!}\int_{X_{0}}^{X}\dx{1} \int_{X_{0}}^{X}\dx{2}\,\mathbb{T}(\TH_{1}\,\TH_{2}) 
\nonumber \\ && \qquad \;
+ \frac{(-\imath)^{3}}{3!}\int_{X_{0}}^{X}\dx{1}\int_{X_{0}}^{X}\dx{2}\int_{X_{0}}^{X}\dx{3}
\,\mathbb{T}(\TH_{1}\,\TH_{2}\,\TH_{3}) 
%%\nonumber \\ && \qquad \;
+ \ldots, \label{eq:Ssum}
\end{eqnarray}
where $\mathbb{T}$ is the space/time-ordering operator. 
Now that it is defined we can simply insert equation (\ref{eq:psiX}) 
into the Schrodinger equation and find that $S$ also obeys the differential
equation 
\begin{equation}
\imath\,\frac{dS}{dx}= \TH\,S
\end{equation}
which describes 9 coupled equations for the elements of $S$. 
From this equation we can also derive that the phase of the determinant $|S| = e^{\imath \Phi}$ 
is simply.
\begin{equation}
\Phi(X,X_{0})=-\int_{X_{0}}^{X}{\rm Tr}(H)\,dx.
\end{equation}
So we find that in the adiabatic basis $\Phi^{(a)}(X,X_{0}) =0$ and does not vary
because the Hamiltonian is traceless. Since 
the initial condition is that $S^{(a)}(X_{0},X_{0})= 1$ we see that in the
adiabatic basis $S$ is a member 
of $SU(3)$. For all bases $S$ has the property that it obeys the product rule
\begin{equation}
S(X,X_{0})=S(X,X_{\star})S(X_{\star},X_{0}). \label{eq:product_rule}
\end{equation}
and the probability that a neutrino with initial state $|\Tnu{j}(X_{0})\rangle$ is
later detected as state $|\Tnu{i}(X)\rangle$ is
\begin{equation}
P(|\Tnu{j}\rangle \rightarrow |\Tnu{i}\rangle) = |S_{ij}(X,X_{0})|^2.
\end{equation}
$S$ is, in general, a member of $U(3)$ and this restriction means 
that any element of $S(X,X_{0})$ satisfies the relationship 
\begin{equation}
S_{ij}(X,X_{0})=e^{\imath \Phi(X,X_{0})}\,C_{ij}^{\ast}(X,X_{0})
\label{eq:Sidentity}
\end{equation}
where $C_{ij}^{\ast}(X,X_{0})$ is the cofactor of the element. This identity
allows us to remove four of 
the elements of $S$ if we know the determinant $|S|$ and there remain two
unitary conditions upon the magnitudes of the 
remaining, independent, elements. Thus in the end we see that $S$, like $U$, is
parameterized by nine real numbers; three 
magnitudes and six phases though the phase of the determinant may be stationary
if the Hamiltonian in that basis is traceless. 

%%%%%%%%%%%%%%%%%%%%%%%%%%%%%%%%%%%%%%%%%%%%%%%%%%%%%%%%%%

\subsection{The two flavor approximation}

We have, so far, described everything in terms of three flavors but a quick
scan through the literature by the reader will reveal that many studies have
used only two.  In this section we show how the three flavor $S$ matrix formalism
using the adiabatic basis can be separated in pieces which contain only two flavors.

The motivation for the reduction in number of flavors one must
consider comes from observation 
of the structure of the adiabatic Hamiltonian, equation (\ref{eq:Ha}). If we make
the assumption that only one $\Gamma_{ij}$ is significant at any given location, never two (or all three) simultaneously,
then neutrino mixing occurs 
only between states $|\Tnu{i}\rangle$ and $|\Tnu{j}\rangle$ and the third state,
$|\Tnu{k}\rangle$, is decoupled. 
It is not immediately obvious 
that only one $\Gamma_{ij}$ is significant at any given location: each
$\Gamma_{ij}$ is proportional to the same derivatives of the potential
$d\dV{\alpha\beta}/dx$ and the only difference is the matter mixing angle
prefactors and the difference between the eigenvalues $\dTk{ij}$ in the
denominators. Nevertheless, this is often the case for most density profiles.

We can express the full Hamiltonian $\TH^{(a)}$ as the sum of three
terms
\begin{subequations}
\begin{eqnarray}
\TH^{(a)} &=&\TH_{(12)}+\TH_{(13)}+\TH_{(23)}\\
&=&
\imath\,\frac{\dTk{12}}{2\,\pi}
\left(\begin{array}{lll} 
0 & \Gamma_{12}\,e^{2\,\imath\pi\dphi{12}} & 0 \\ 
-\Gamma^{\ast}_{12}\,e^{-2\,\imath\pi\dphi{12}} & 0 & 0 \\ 
0 & 0 & 0 \end{array}\right)
%%\nonumber \\ &&
+\imath\,\frac{\dTk{13}}{2\,\pi}\left(\begin{array}{lll} 
0 & 0 & \Gamma_{13}\,e^{2\,\imath\pi\dphi{13}} \\ 
0 & 0 & 0 \\ 
-\Gamma^{\ast}_{13}\,e^{-2\,\imath\pi\dphi{13}} & 0 & 0 
\end{array}\right)
\nonumber \\ && \;
+\imath\,\frac{\dTk{23}}{2\,\pi}\left(\begin{array}{lll} 
0 & 0 & 0 \\ 
0 & 0 & \Gamma_{23}\,e^{2\,\imath\pi\dphi{23}} \\ 
0 & -\Gamma^{\ast}_{23}\,e^{-2\,\imath\pi\dphi{23}} & 0 
\end{array}\right).
%%\nonumber \\
\end{eqnarray} 
\end{subequations}
and then introduce the three $S$-matrices $S_{(12)}$, $S_{(13)}$ and $S_{(23)}$
we obtain from the substitution of each $H_{(ij)}$ into equation
(\ref{eq:Ssum}). 
Due to the structure of each $H_{(ij)}$ the three $S_{(ij)}$ have the form
\begin{subequations}
\begin{eqnarray}
S_{(12)}&=&\left(\begin{array}{lll} 
\zeta_{(12)} & \eta_{(12)} & 0 \\ 
-\eta_{(12)}^{\ast} & \zeta_{(12)}^{\ast} & 0 \\ 
0 & 0 & 1 
\end{array}\right)\\
S_{(13)}&=&\left(\begin{array}{lll} 
\zeta_{(13)} & 0 & \eta_{(13)} \\ 
0 & 1 & 0 \\ 
-\eta_{(13)}^{\ast} & 0 & \zeta_{(13)}^{\ast} 
\end{array}\right)\\
S_{(23)}&=&\left(\begin{array}{lll} 
1 & 0 & 0 \\ 
0 & \zeta_{(23)} & \eta_{(23)} \\ 
0 & -\eta_{(23)}^{\ast} & \zeta_{(23)}^{\ast} 
\end{array}\right)
\end{eqnarray} 
\end{subequations}
where the $\zeta_{(ij)}$'s and $\eta_{(ij)}$'s are Cayley Klein parameters. 
Now it is simply a case of adapting this approximation to the situation at
hand. 
In figure (\ref{fig:k1k2k3}) we see that the L resonance always involves mixing
between
states $|\Tnu{1}\rangle$ and $|\Tnu{2}\rangle$ so if the density profile under
consideration 
possesses only an L resonance then the $S$ matrix describing the evolution of
the neutrinos through 
the profile will have the structure of $S_{(12)}$. The H resonance mixes states
$|\Tnu{2}\rangle$ and $|\Tnu{3}\rangle$
for a normal hierarchy so for a profile containing just an H resonance the
$S$ matrix will have the
structure of $S_{(23)}$. By applying the same reasoning for all the different
possibilities we can assign 
the appropriate $S_{(ij)}$ for any resonance shown in the figure. 
For a density profile possessing multiple resonances we can apply the group
product rule for the evolution 
operator, equation (\ref{eq:product_rule}), to divide the profile into
sub-domains such that, within each, there is 
just one resonance. Since the structure of the $S$ matrix for each sub-domain is
given by the reasoning above the $S$ matrix 
for the entire profile is then the time-ordered product of the appropriate
$S_{(ij)}$'s.

Once we have decided which pair of states are mixing we can then either solve
the reduced problem 
\begin{equation}
\imath\,\frac{dS^{(2)}}{dx}= \TH_{(ij)}^{(2)}\,S^{(2)} 
\end{equation} 
where $S^{(2)}$ is a 2x2 matrix and 
\begin{equation}
\TH_{(ij)}^{(2)}=\imath\,\frac{\dTk{ij}}{2\,\pi} \left(\begin{array}{cc} 0 &
\Gamma_{ij}\,e^{2\,\imath\pi\dphi{ij}} \\ 
-\Gamma^{\ast}_{ij}\,e^{-2\,\imath\pi\dphi{ij}} & 0
\end{array}\right).
\end{equation} 
using the expression for $\Gamma_{ij}$ given in equations
(\ref{eq:Gamma12})-(\ref{eq:Gamma23}), 
use a straight two flavor calculation a la Kneller and McLaughlin
\cite{K&M2006}, or utilize 
some other alternative or approximate method. However one determines $S^{(2)}$,
once it has been found 
one simply constructs the appropriate three flavor $S$ matrix (or matrices in
the case profiles with L, H and/or $\mu\tau$ resonances) as in, for example,
Kneller, McLaughlin \& Brockman \cite{2008PhRvD..77d5023K}. The disadvantage
of applying a series of two flavor approximations is that some phase information can be lost. 
Nevertheless, many features of a flavor transformed neutrino signal can often be determined
in this way.

%%%%%%%%%%%%%%%%%%%%%%%%%%%%%%%%%%%%%%%%%%%%%%%%%%%%%%%%%%%%%%%%%%%%%%%%

\subsection{Identities of the Scattering Matrix} 
\label{sec:identities}

The problem of neutrino propagation through supernovae has received
considerable 
attention the in the past few years. The signal from the next Galactic supernova
has the potential to reveal a great deal of information about both the supernova
and the mixing parameters for the neutrinos. For example, if the angle $\theta_{13}$
is not too small 
then dynamic MSW effects may be observed and some authors have also 
considered the possibility of observing effects from a non-zero CP phase. It is
upon this 
possibility of observing the non-zero CP phase that we now focus our attention. 

In the adiabatic and matter basis we have shown that the Hamiltonian is
independent 
of the CP phase $\epsilon$ if $V_\mu=V_\tau=0$. This occurs because the
eigenvalues 
and $\Tt{12}$ and $\Tt{13}$ are all independent of $\epsilon$ in this limit so
that the 
non-adiabaticity parameters $\Gamma_{12}$, $\Gamma_{13}$ and $\Gamma_{23}$ are
also independent of $\epsilon$. If 
the Hamiltonian is independent of $\epsilon$ then the $S$ matrix must also be
independent 
of $\epsilon$ and, therefore, we must have $S^{(a)}(X,X_0,\epsilon) =
S^{(a)}(X,X_0,0)$ and $S^{(m)}(X,X_0,\epsilon) =
S^{(m)}(X,X_0,0)$. 
As a result, all the survival and crossing probabilities for the matter or
adiabatic neutrino states,
$P(|\Tnu{j}\rangle\rightarrow|\Tnu{i}\rangle)=|S^{(m,a)}_{ij}|^{2}$, are also
independent of $\epsilon$. 
Any dependence upon $\epsilon$ for the survival/crossing probabilities of states
in other bases 
can only enter explicitly in the unitary transformation to those bases. 

For example, let us consider the transformation to the flavor basis. 
The flavor basis survival/crossing probabilities are found from the $S^{(f)}$
matrix related to 
$S^{(m)}(X,X_{0},\epsilon)$ and $S^{(a)}(X,X_0,\epsilon)$ by
\begin{eqnarray}
&& 
S^{(f)}(X,X_{0},\epsilon)
%%\nonumber \\ && \qquad 
=\TU(X,\epsilon)\,S^{(m)}(X,X_{0},\epsilon)\,\TU^{\dagger}(X_{0},\epsilon) 
%%\nonumber \\ && \qquad
=\TU(X,\epsilon)\,W(X,\epsilon)\,S^{(a)}(X,X_{0},\epsilon)
%%\nonumber \\ && \qquad\qquad\times 
W^\dagger(X_0,\epsilon)\TU^{\dagger}(X_{0},\epsilon).
%%\nonumber \\ &&
\end{eqnarray}
In above expression we can set $W^\dagger(X_0,\epsilon) = 1$  if all the phases 
in equation (\ref{eq:W}) are set to zero at $X_0$. However
when splitting the $S$ matrix into pieces, as in equation (\ref{eq:product_rule}), then one should
retain this phase information, so $W^\dagger = 1$ only for the first, rightmost, $S$ matrix.

Since $S^{(m)}(X,X_{0},\epsilon) = S^{(m)}(X,X_{0},0)$ the following
identity must be obeyed:
\begin{eqnarray}
&&
\TU^{\dagger}(X,\epsilon)\,S^{(f)}(X,X_{0},\epsilon)\,\TU(X_{0},\epsilon)
%%\nonumber \\&&\qquad
=\TU^{\dagger}(X,0)\,S^{(f)}(X,X_{0},0)\,\TU(X_{0},\epsilon=0). 
%%\nonumber \\ &&
\label{eq:Sfidentity}
\end{eqnarray}
Now let us consider how the CP phase enters into $\TU$. We saw from equation 
(\ref{eq:tildeUdecomposition}) that $\TU$ could be written as
$\TU =
B(\Tb,\Td)\,\Theta_{23}(\Tt{23})\,E(\Te)\,\Theta_{13}(\Tt{13})\,\Theta_{12}(\Tt{12})\,
A(\Ta{1},\Ta{2},\Ta{3})$ but, when $V_\mu=V_\tau=0$, we have also seen that both $\Tt{12}$ and $\Tt{13}$
are independent of $\epsilon$ and, furthermore, we have also made the decision that $\Ta{1}$, $\Ta{2}$ and $\Ta{3}$
are constants. 
This means $\Theta_{12}(x,\epsilon)=\Theta_{12}(x,0)$,
$\Theta_{13}(x,\epsilon)=\Theta_{13}(x,0)$ and
$A(x,\epsilon)=A(x,0)$ so when we insert the decomposition of $\TU$ 
into equation (\ref{eq:Sfidentity}) we find that what survives can be written as 
\begin{eqnarray}
&&
B^{\dagger}(X,\epsilon)\,S^{(f)}(X,X_{0},\epsilon)\,B(X_{0},\epsilon)
\nonumber \\ && \qquad 
=\Theta_{23}(X,\epsilon)\,E(X,\epsilon)\,\Theta_{23}^{\dagger}(X,0)
%%\nonumber \\ && \qquad\qquad\times 
B^{\dagger}(X,0)\,S^{(f)}(X,X_{0},0)\,B(X_{0},0) 
%%\nonumber \\ && \qquad\qquad\times 
\Theta_{23}(X_{0},0)\,E^{\dagger}(X_{0},\epsilon)\,\Theta^{\dagger}_{23}(X_{0},\epsilon).\label{eq:Sfidentity2}
\end{eqnarray}
The two combinations $\Theta_{23}(X,\epsilon)\,E(X,\epsilon)\,\Theta_{23}^{\dagger}(X,0)$ 
and
$\Theta_{23}(X_{0},0)\,E^{\dagger}(X_{0},\epsilon)\,\Theta^{\dagger}_{23}(X_{0},\epsilon)$
are
\begin{widetext}
\begin{eqnarray}
&& \Theta_{23}(X,\epsilon)\,E(X,\epsilon)\,\Theta_{23}^{\dagger}(X,0) =  \nonumber \\
&& \qquad\;\left(\begin{array}{lll} 1 & 0 & 0 \\ 0 &
\Ts{23}(X,\epsilon)\Ts{23}(X,0)+\Tc{23}(X,\epsilon)\Tc{23}(X,0)e^{\iTe(X)} &
\Ts{23}(X,\epsilon)\Tc{23}(X,0)-\Tc{23}(X,\epsilon)\Ts{23}(X,0)e^{\iTe(X)} \\ 0
& \Tc{23}(X,\epsilon)\Ts{23}(X,0)-\Ts{23}(X,\epsilon)\Tc{23}(X,0)e^{\iTe(X)} &
\Tc{23}(X,\epsilon)\Tc{23}(X,0)+\Ts{23}(X,\epsilon)\Ts{23}(X,0)e^{\iTe(X)}
\end{array}\right) ,\\
&&
\Theta_{23}(X_{0},0)\,E^{\dagger}(X_{0},\epsilon)\,\Theta^{\dagger}_{23}(X_{0},
\epsilon)=
\nonumber \\ && \qquad\;
\left(\begin{array}{lll} 1 & 0 & 0 \\ 0 &
\Ts{23}(X_{0},\epsilon)\Ts{23}(X_{0},0)+\Tc{23}(X_{0},\epsilon)\Tc{23}(X_{0},
0)e^{-\iTe(X_{0})} &
\Tc{23}(X_{0},\epsilon)\Ts{23}(X_{0},0)-\Ts{23}(X_{0},\epsilon)\Tc{23}(X_{0},
0)e^{-\iTe(X_{0})} \\ 0 &
\Ts{23}(X_{0},\epsilon)\Tc{23}(X_{0},0)-\Tc{23}(X_{0},\epsilon)\Ts{23}(X_{0},
0)e^{-\iTe(X_{0})} &
\Tc{23}(X_{0},\epsilon)\Tc{23}(X_{0},0)+\Ts{23}(X_{0},\epsilon)\Ts{23}(X_{0},
0)e^{-\iTe(X_{0})} \end{array}\right)
\nonumber \\
\end{eqnarray}
\end{widetext}
respectively.
After inserting these matrices into equation (\ref{eq:Sfidentity2}) we find nine
relationships 
between the elements of $S^{(f)}(X,X_{0},\epsilon)$ and  
the elements of $S^{(f)}(X,X_{0},0)$. The flavour basis
survival/crossing 
probabilities $P_{\alpha\beta}=|S^{(f)}_{\alpha\beta}|^{2}$ may be found for both
$S^{(f)}(X,X_{0},\epsilon)$ and 
$S^{(f)}(X,X_{0},0)$ and using the relationships between the elements
of the two matrices 
we find that we can derive four, non-trivial identities for the probabilities:
\begin{itemize}
 \item $P_{ee}(X,X_{0},\epsilon) = P_{ee}(X,X_{0},0)$,
 \item $P_{e\mu}(X,X_{0},\epsilon)+P_{e\tau}(X,X_{0},\epsilon) =
P_{e\mu}(X,X_{0},0)+P_{e\tau}(X,X_{0},0)$,
 \item $P_{\mu e}(X,X_{0},\epsilon)+P_{\tau e}(X,X_{0},\epsilon) =
P_{\mu e}(X,X_{0},0)+P^{(f)}_{\tau e}(X,X_{0},0)$,
 \item
$P_{\mu\mu}(X,X_{0},\epsilon)+P_{\mu\tau}(X,X_{0},\epsilon)+P_
{\tau\mu}(X,X_{0},\epsilon)+P_{\tau\tau}(X,X_{0},\epsilon)=P_{\mu\mu}(X,X_{0},0)+P_{\mu\tau}(X,X_{0},0)+P_{\tau\mu}(X,X_
{0},0)+P_{\tau\tau}(X,X_{0},0)$.
\end{itemize}
The first two identities were also found by Balantekin, Gava and Volpe \cite{Balantekin:2007es}. 

However when computing the fluxes at Earth one requires a slightly different set
of probabilities: the probability that an initial \emph{flavor} state emerges as a
given \emph{matter/mass} state. These probabilities are found from the matrix $S^{(mf)}$ given by 
\begin{eqnarray}
&&S^{(mf)}(X,X_{0},\epsilon)=S^{(m)}(X,X_{0},\epsilon)\,\TU^{\dagger}(X_{0},\epsilon)
%%\nonumber \\ &&\qquad
=W(X,\epsilon)\,S^{(a)}(X,X_{0},\epsilon)\, W^\dagger(X_{0},\epsilon)  \TU^{\dagger}(X_{0},\epsilon).
\nonumber \\
\end{eqnarray}
Again the invariance of $S^{(m)}(X,X_{0},\epsilon)$ with regard to $\epsilon$
when $V_\mu=V_\tau=0$ means that we have the following identity:
\begin{eqnarray}
&&
S^{(mf)}(X,X_{0},\epsilon)\,\TU(X_{0},\epsilon) 
%%\nonumber \\&&\qquad
=S^{(mf)}(X,X_{0},0)\,\TU(X_{0},0). \label{eq:Smfidentity}
\end{eqnarray}
which ultimately leads to 
\begin{eqnarray}
&&
S^{(mf)}(X,X_{0},\epsilon)\,B(X_{0},\epsilon) 
%%\nonumber \\&&\qquad
=S^{(mf)}(X,X_{0},0)\,B(X_{0},0)\,
%%\nonumber \\&&\qquad\qquad\times
\Theta_{23}(X_{0},0)\,E^{\dagger}(X_{0},\epsilon)\,\Theta^{\dagger}_{23}(X_{0},\epsilon).
\end{eqnarray}
This is very similar to equation (\ref{eq:Sfidentity2}). 
Again we have nine relationships between $S^{(mf)}(X,X_{0},\epsilon)$ and
$S^{(mf)}(X,X_{0},0)$ which lead to the following, non-trivial identities
for the probabilities $P_{i\alpha}=|S^{(mf)}_{i\alpha}|^{2}$:
\begin{itemize}
 \item $P_{1e}(X,X_{0},\epsilon) = P_{1e}(X,X_{0},0)$,
 \item $P_{2e}(X,X_{0},\epsilon)= P_{2e}(X,X_{0},0)$,
 \item $P_{3e}(X,X_{0},\epsilon)= P_{3e}(X,X_{0},0)$,  
 \item $P_{1\mu}(X,X_{0},\epsilon)+P_{1\tau}(X,X_{0},\epsilon) = P_{1\mu}(X,X_{0},0)+P_{1\tau}(X,X_{0},0)$,
 \item $P_{2\mu}(X,X_{0},\epsilon)+P_{2\tau}(X,X_{0},\epsilon) =P_{2\mu}(X,X_{0},0)+P_{2\tau}(X,X_{0},0)$,
 \item $P_{3\mu}(X,X_{0},\epsilon)+P_{3\tau}(X,X_{0},\epsilon)=P_{3\mu}(X,X_{0},0)+P_{3\tau}(X,X_{0},0)$.
\end{itemize}
At the detector on Earth the supernova neutrinos are measured as flavor eigenstates thus, for example, the electron neutrino
signal is the appropriate linear combination of survival probabilities multiplied but the original fluxes.  
One can see from these identities above that the CP phase effects will not
show up in observed supernova neutrino signal if the $\nu_\mu, \nu_\tau$ fluxes emitted from the supernova 
neutrino sphere are equal \cite{Akhmedov:2002zj,Balantekin:2007es}.
However, as pointed out in \cite{Balantekin:2007es} if the $\nu_\mu, \nu_\tau$ fluxes are not equal
when they are emitted, then there is no guarantee that the fluxes as observed in the detector will be independent of the CP phase.

In the case where when $\epsilon\neq 0$ \emph{and} $\delta V_{\mu\tau}\neq 0$ these identities
no longer apply because the eigenvalues and the mixing angles now become 
functions of $\epsilon$.  One might imagine that some sort of cancellation occurs
such that the three non-adiabaticity parameters are independent of $\epsilon$. 
However an examination of the adiabatic basis shows that a complete
cancellation cannot occur.
For all three non-adiabaticity parameters, equations (\ref{eq:Gamma12})-(\ref{eq:Gamma23}),
there emerges an imaginary component proportional to 
$\sin\epsilon\,d\dV{\mu\tau}/dx$ and this term cannot be canceled
by concomitant changes in the eigenvalues and/or mixing angles. For all 
three $\Gamma_{ij}$ we have a situation where $\Gamma_{ij}(\epsilon) \neq\Gamma_{ij}(0)$ thus 
$\TH^{(a)}(\epsilon) \neq \TH^{(a)}(0)$, $S^{(a)}(\epsilon) \neq S^{(a)}(0)$ and,
finally, $P_{ij}(\epsilon) \neq P_{ij}(0)$ for any basis. 

%%%%%%%%%%%%%%%%%%%%%%%%%%%%%%%%%%%%%%%%%%%%%%%%%%%%%%%%%%

\section{Summary And Conclusions} \label{sec:conclusions}

We have considered the problem of a generalized, $3$-flavor, neutrino
mixing that includes matters potentials for both $\mu$ and $\tau$ flavors and CP violation. We
presented expressions for the eigenvalues and matter mixing angles and 
pointed out that mixing phases that were zeroed in the vacuum are not necessarily zero in matter. 
We found that in the limit that the mu and tau potentials are equal,
the eigenvalues and matter mixing angles $\tilde{\theta}_{12}$ and
$\tilde{\theta}_{13}$ 
are independent of the CP phase, although $\tilde{\theta}_{23}$ does have 
CP dependence.

We introduced the 3-flavor adiabatic basis.  In this basis the 
Hamiltonian is completely off-diagonal and the behavior of the neutrinos is largely
determined by the $3$-flavor adiabaticity parameters. This is a useful basis in which to calculate
neutrino flavor transformation; further it gives a straightforward picture of the
effects of the CP phase.   In the limit
that the mu and tau neutrino potentials are the same, the Hamiltonian in the adiabatic basis 
is independent of both $\tilde{\theta}_{23}$ and the CP violating phase, so CP phase effects 
appear only in rotations into and out of the flavor basis.

We discussed the $S$ matrix for three flavors, as well as two flavor S-matrix 
approximation.  Using the three flavor S-matrix, we found 
several non-trivial identities related to the observability of the CP phase. In a 
future study \cite{K&M2009} we shall discuss how one can formulate efficient algorithms for the calculation of $S$ and present some calculations of three flavor oscillations made with them.

%%%%%%%%%%%%%%%%%%%%%%%%%%%%%%%%%%%%%%%%%%%%%%%%%%%%%%%%%%

\begin{acknowledgments}
The authors are grateful to Baha Balantekin, Jerome Gava, and Cristina
Volpe for 
their suggestions while we were preparing this paper. 
This work was supported in part by US DOE grant DE-FG02-87ER40328 at UMN, and DE-FG02-02ER41216 at
NC State and by
``Non standard 
neutrino properties and their impact in astrophysics and cosmology'', Project
No.\ ANR-05-JCJC-0023 at IPN Orsay. 
\end{acknowledgments}

%%%%%%%%%%%%%%%%%%%%%%%%%%%%%%%%%%%%%%%%%%%%%%%%%%%%%%%%%%
%%%%%%%%%%%%%%%%%%%%%%%%%%%%%%%%%%%%%%%%%%%%%%%%%%%%%%%%%%
%%%%%%%%%%%%%%%%%%%%%%%%%%%%%%%%%%%%%%%%%%%%%%%%%%%%%%%%%%

%%%%%%%%%%%%%%%%%%%%%%%%%%%%%%%%%%%%%%%%%%%%%%%%%%%%%%%%%%


\begin{thebibliography}{99}

\bibitem{P1957} B.~Pontecorvo, Sov. Phys. JETP, {\bf 33}, 549 (1957)

\bibitem{P1958} B.~Pontecorvo Sov. Phys. JETP, {\bf 34}, 247 (1958)

\bibitem{Maki:1962mu} Z.~Maki, M.~Nakagawa and S.~Sakata, Prog. Theor. Phys., {\bf 28}, 870 (1962)

\bibitem{Nakagawa:1963uw} M.~Nakagawa, H.~Okonogi, S.~Sakata and A.~Toyoda, Prog. Theor. Phys., {\bf 30}, 727 (1963)

\bibitem{M&S1986} S.~P.~Mikheev and A.~I.~Smirnov, Nuovo Cimento C, {\bf 9}, 17, (1986)

\bibitem{Wolfenstein1977} L.~Wolfenstein, Phys. Rev., {\bf D17}, 2369 (1978)

\bibitem{PDG2006} W.-M.Yao et al. (Particle Data Group), J. Phys. G, {\bf 33}, 1 (2006) 

\bibitem{2006PrPNP..57..742F} G.~L.~Fogli, E.~Lisi, A.~Marrone and A.~Palazzo, Progress in Particle and Nuclear Physics, {\bf 57}, 742 (2006)

\bibitem{2001PhRvL..87g1301A} Ahmad, Q.~R., et al., \prl, {\bf 87}, 071301 (2001)

\bibitem{Schirato:2002tg} R.~C.~Schirato and G.~M.~Fuller, arXiv:astro-ph/0205390

\bibitem{Takahashi:2002yj} K.~Takahashi, K.~Sato, H.~E.~Dalhed and J.~R.~Wilson, Astropart. Phys., {\bf 20}, 189 (2003)

\bibitem{Fogli:2003dw} Fogli, G.~L., Lisi, E., Mirizzi, A. and Montanino, D., \prd, {\bf 68}, 033005 (2003)

\bibitem{Tomas:2004gr} R.~Tomas, M.~Kachelriess, G.~Raffelt, A.~Dighe, H.~T.~Janka and L.~Scheck, JCAP, {\bf 0409}, 015 (2004)

%\bibitem{Fogli:2006xy} G.~L.~Fogli, E.~Lisi, A.~Mirizzi and D.~Montanino, JCAP, {\bf 0606}, 012 (2006)

\bibitem{Dasgupta:2005wn} B.~Dasgupta and A.~Dighe, \prd, {\bf 75}, 093002 (2007)

\bibitem{Friedland:2006ta} A.~Friedland and A.~Gruzinov, arXiv:astro-ph/0607244.

\bibitem{K&M2006} J.~P.~Kneller and G.~C.~McLaughlin, \prd, {\bf 73}, 056003 (2006)

%\cite{Fogli:2006xy}
\bibitem{Fogli:2006xy}
  G.~L.~Fogli, E.~Lisi, A.~Mirizzi and D.~Montanino,
  %``Damping of supernova neutrino transitions in stochastic shock-wave density profiles,''
  JCAP, {\bf 0606}, 012 (2006)
  [arXiv:hep-ph/0603033].
  %%CITATION = JCAPA,0606,012;%%

%\cite{Choubey:2007ga}
\bibitem{Choubey:2007ga}
  S.~Choubey, N.~P.~Harries and G.~G.~Ross,
  %``Turbulent supernova shock waves and the sterile neutrino signature in megaton water detectors,''
  \prd, {\bf 76}, 073013 (2007)
  [arXiv:hep-ph/0703092].
  %%CITATION = PHRVA,D76,073013;%%

\bibitem{2008PhRvD..77d5023K} J.~P.~Kneller, G.~C.~McLaughlin and J.~Brockman, \prd, {\bf 77}, 045023 (2008)

\bibitem{Pantaleone:1992eq}
  J.T.~Pantaleone,
%  ``Neutrino oscillations at high densities,''
  Phys.\ Lett.\  B, {\bf 287}, 128 (1992).
  %%CITATION = PHLTA,B287,128;%%

\bibitem{Samuel:1993uw}
  S.~Samuel,
%  ``Neutrino oscillations in dense neutrino gases,''
  \prd, {\bf 48}, 1462 (1993).
  %%CITATION = PHRVA,D48,1462;%%

\bibitem{Qian:1995ua}
  Y.Z.~Qian and G.M.~Fuller,
%  ``Matter enhanced anti-neutrino flavor transformation and supernova nucleosynthesis,''
  \prd, {\bf 52}, 656 (1995)
  [arXiv:astro-ph/9502080].
  %%CITATION = PHRVA,D52,656;%%

\bibitem{Pastor:2002we}
  S.~Pastor and G.~Raffelt,
%  ``Flavor oscillations in the supernova hot bubble region: Nonlinear effects of neutrino background,''
  \prl, {\bf 89}, 191101 (2002)
  [arXiv:astro-ph/0207281].
  %%CITATION = PRLTA,89,191101;%%

%\cite{Balantekin:2004ug}
\bibitem{Balantekin:2004ug}
  A.~B.~Balantekin and H.~Yuksel,
  %``Neutrino mixing and nucleosynthesis in core-collapse supernovae,''
  New J.\ Phys.,  {\bf 7}, 51 (2005)
  [arXiv:astro-ph/0411159].
  %%CITATION = NJOPF,7,51;%%

\bibitem{Sawyer:2005jk}
  R.F.~Sawyer,
%  ``Speed-up of neutrino transformations in a supernova environment,''
  \prd, {\bf 72}, 045003 (2005)
  [arXiv:hep-ph/0503013].
  %%CITATION = PHRVA,D72,045003;%%

\bibitem{Fuller:2005ae}
  G.M.~Fuller and Y.Z.~Qian,
%  ``Simultaneous flavor transformation of neutrinos and antineutrinos
%  with dominant potentials from neutrino neutrino forward scattering,''
  \prd, {\bf 73}, 023004 (2006)
  [arXiv:astro-ph/0505240].
  %%CITATION = PHRVA,D73,023004;%%

\bibitem{Duan:2005cp}
  H.~Duan, G.M.~Fuller and Y.Z.~Qian,
%  ``Collective neutrino flavor transformation in supernovae,''
  \prd, {\bf 74}, 123004 (2006)
  [arXiv:astro-ph/0511275].
  %%CITATION = PHRVA,D74,123004;%%

\bibitem{Duan:2006an} 
H.~Duan, G.M.~Fuller, J.~Carlson and Y.Z.~Qian,
%    ``Simulation of coherent non-linear neutrino flavor transformation in the supernova environment. I: Correlated
%    neutrino trajectories,'' 
\prd, {\bf 74}, 105014 (2006)
    [arXiv:astro-ph/0606616].
%%CITATION = ASTRO-PH 0606616;%%

\bibitem{Hannestad:2006nj}
  S.~Hannestad, G.G.~Raffelt, G.~Sigl and Y.Y.Y.~Wong,
%  ``Self-induced conversion in dense neutrino gases:
%  Pendulum in flavor space,''
  \prd, {\bf 74}, 105010 (2006); {\bf 76}, 029901(E) (2007)
%  Erratum ibid.\ {\bf 76}, 029901 (2007).
  [arXiv:astro-ph/0608695].
%%CITATION = ASTRO-PH 0608695;%%

\bibitem{Balantekin:2006tg}
  A.B.~Balantekin and Y.~Pehlivan,
  %``Neutrino neutrino interactions and flavor mixing in dense matter,''
  J.\ Phys.\ G, {\bf 34}, 47 (2007)
  [arXiv:astro-ph/0607527].
  %%CITATION = JPHGB,G34,47;%%

\bibitem{Duan:2007mv}
  H.~Duan, G.M.~Fuller, J.~Carlson and Y.Z.~Qian,
%  ``Analysis of collective neutrino flavor transformation in supernovae,''
  \prd, {\bf 75}, 125005 (2007)
  [arXiv:astro-ph/0703776].
%%CITATION = PHRVA,D75,125005;%%

\bibitem{Raffelt:2007yz}
  G.G.~Raffelt and G.~Sigl,
%  ``Self-induced decoherence in dense neutrino gases,''
  \prd, {\bf 75}, 083002 (2007)
  [arXiv:hep-ph/0701182]
%%CITATION = HEP-PH/0701182;%%

\bibitem{EstebanPretel:2007ec}
 A.~Esteban-Pretel, S.~Pastor, R.~Tom\`as,  G.G.~Raffelt and G.~Sigl,
% ``Decoherence in supernova neutrino transformations suppressed by deleptonization,''
 \prd, {\bf 76}, 125018 (2007)
 [arXiv:0706.2498].
 %%CITATION = ARXIV:0706.2498;%%

\bibitem{Raffelt:2007cb}
  G.G.~Raffelt and A.Yu.~Smirnov,
%  ``Self-induced spectral splits in supernova neutrino fluxes,''
  \prd, {\bf 76}, 081301 (2007)
  [arXiv:0705.1830].
%%CITATION = PHRVA,D76,081301;%%

\bibitem{Raffelt:2007xt}
 G.G.~Raffelt and A.Yu.~Smirnov,
% ``Adiabaticity and spectral splits in collective neutrino transformations,''
 \prd, {\bf 76}, 125008 (2007)
 [arXiv:0709.4641].
%%CITATION = ARXIV:0709.4641;%%

\bibitem{Duan:2007fw}
 H.~Duan, G.M.~Fuller and Y.Z.~Qian,
% ``A simple picture for neutrino flavor transformation in supernovae,''
 \prd, {\bf 76}, 085013 (2007)
 [arXiv:0706.4293].
%%CITATION = ARXIV:0706.4293;%%

\bibitem{Fogli:2007bk}
 G.L.~Fogli, E.~Lisi, A.~Marrone and A.~Mirizzi,
% ``Collective neutrino flavor transitions in supernovae and the role of trajectory averaging,''
  J.\ Cosmol.\ Astropart.\ Phys.\ {\bf 0712}, 010 (2007)
%JCAP {\bf 0712}, 010 (2007)
 [arXiv:0707.1998].
%%CITATION = ARXIV:0707.1998;%%

\bibitem{Duan:2007bt}
 H.~Duan, G.M.~Fuller, J.~Carlson and Y.Z.~Qian,
% ``Neutrino mass hierarchy and stepwise spectral swapping of supernova
% neutrino flavors,''
 \prl, {\bf 99}, 241802 (2007)
 [arXiv:0707.0290].
%%CITATION = ARXIV:0707.0290;%%

\bibitem{Duan:2007sh}
 H.~Duan, G.M.~Fuller, J.~Carlson and Y.Z.~Qian,
 %``Flavor evolution of the neutronization neutrino burst from an O-Ne-Mg core-collapse supernova,''
  \prl  {\bf 100}, 021101 (2008)
  [arXiv:0710.1271].
  %%CITATION = PRLTA,100,021101;%%

%\cite{EstebanPretel:2007yu}
\bibitem{EstebanPretel:2007yu}
  A.~Esteban-Pretel, R.~Tomas and J.~W.~F.~Valle,
  %``Probing non-standard neutrino interactions with supernova neutrinos,''
  \prd, {\bf 76}, 053001 (2007)
  [arXiv:0704.0032 [hep-ph]].
  %%CITATION = PHRVA,D76,053001;%%

\bibitem{Dasgupta:2008cd}
  B.~Dasgupta, A.~Dighe, A.~Mirizzi and G.G.~Raffelt,
%  ``Spectral split in prompt supernova neutrino burst: Analytic three-flavor treatment,'' 
  \prd, {\bf 77}, 113007 (2008)
  [arXiv:0801.1660].
  %%CITATION = ARXIV:0801.1660;%%

\bibitem{EstebanPretel:2007yq}
  A.~Esteban-Pretel, S.~Pastor, R.~Tom\`as, G.G.~Raffelt and G.~Sigl,
%  ``Mu-tau neutrino refraction and collective three-flavor
%  transformations in supernovae,''
  \prd, {\bf 77}, 065024 (2008)
  [arXiv:0712.1137].
  %%CITATION = PHRVA,D77,065024;%%

%\cite{Dasgupta:2007ws}
\bibitem{Dasgupta:2007ws}
  B.~Dasgupta and A.~Dighe,
  %``Collective three-flavor oscillations of supernova neutrinos,''
  \prd, {\bf 77}, 113002 (2008)
  [arXiv:0712.3798 [hep-ph]].
  %%CITATION = PHRVA,D77,113002;%%

%\cite{Duan:2008za}              }
\bibitem{Duan:2008za}
  H.~Duan, G.M.~Fuller and Y.Z.~Qian,
%  ``Stepwise spectral swapping with three neutrino flavors,''
   \prd, {\bf 77}, 085016 (2008)
  [arXiv:0801.1363].
  %%CITATION = ARXIV:0801.1363;%%

\bibitem{Dasgupta:2008my}
  B.~Dasgupta, A.~Dighe and A.~Mirizzi,
%  ``Identifying neutrino mass hierarchy at extremely small $\Theta_{13}$ through Earth matter effects in a supernova signal,''
\prl, {\bf 101}, 171801 (2008)
[arXiv:0802.1481].
  %%CITATION = ARXIV:0802.1481;%%

\bibitem{Sawyer:2008zs}
  R.F.~Sawyer,
%  ``The multi-angle instability in dense neutrino systems,''
  preprint arXiv:0803.4319.
  %%CITATION = ARXIV:0803.4319;%%

%\cite{Duan:2008eb}
\bibitem{Duan:2008eb}
  H.~Duan, G.M.~Fuller and J.~Carlson,
%  ``Simulating nonlinear neutrino flavor evolution,''
  Computational Science and Discovery, {\bf 1}, 015007 (2008)
  %%CITATION = ARXIV:0803.3650;%%

\bibitem{Chakraborty:2008zp}
S.~Chakraborty, S.~Choubey, B.~Dasgupta and K.~Kar,
%  ``Effect of collective flavor oscillations on the diffuse supernova neutrino background,''
J.\ Cosmol.\ Astropart.\ Phys.\ {\bf 0809}, 013 (2008)
 [arXiv:arXiv:0805.3131].
  %%CITATION = ARXIV:0805.3131;%%

%\cite{Dasgupta:2008cu}
\bibitem{Dasgupta:2008cu}
  B.~Dasgupta, A.~Dighe, A.~Mirizzi and G.G.~Raffelt,
%  ``Collective neutrino oscillations in non-spherical geometry,''
  \prd, {\bf 78}, 033014 (2008)
 [arXiv:0805.3300].
  %%CITATION = ARXIV:0805.3300;%%

%\cite{EstebanPretel:2008ni}
\bibitem{EstebanPretel:2008ni}
  A.~Esteban-Pretel, A.~Mirizzi, S.~Pastor, R.~Tomas, G.~G.~Raffelt, P.~D.~Serpico and G.~Sigl,
  %``Role of dense matter in collective supernova neutrino transformations,''
  \prd, {\bf 78}, 085012 (2008)
  [arXiv:0807.0659 [astro-ph]].
  %%CITATION = PHRVA,D78,085012;%%

%\cite{Sigl:2009cw}
\bibitem{Sigl:2009cw}
  G.~Sigl, R.~Tomas, A.~Esteban-Pretel, S.~Pastor, A.~Mirizzi, G.~G.~Raffelt and P.~D.~Serpico,
  %``Collective flavor transitions of supernova neutrinos,''
  arXiv:0901.0725 [hep-ph].
  %%CITATION = ARXIV:0901.0725;%%

\bibitem{EstebanPretel:2007bz} A.~Esteban-Pretel, S.~Pastor, R.~Tomas, G.~G.~Raffelt and G.~Sigl, arXiv:0712.2176

\bibitem{Botella:1987aa} Botella, F. J. Lim, C. -S. and Marciano, W. J., \prd, {\bf 35} 896 (1987)

%\cite{Gava:2009pj}
\bibitem{Gava:2009pj}
  J.~Gava, J.~Kneller, C.~Volpe and G.~C.~McLaughlin,
  %``A dynamical collective calculation of supernova neutrino signals,''
  arXiv:0902.0317 [hep-ph].
  %%CITATION = ARXIV:0902.0317;%%

%\cite{Akhmedov:2002zj}
\bibitem{Akhmedov:2002zj}
  E.~K.~Akhmedov, C.~Lunardini and A.~Y.~Smirnov,
  %``Supernova neutrinos: Difference of nu/mu - nu/tau fluxes and conversion effects,''
  Nucl.\ Phys.\  B, {\bf 643}, 339 (2002)
  [arXiv:hep-ph/0204091].
  %%CITATION = NUPHA,B643,339;%%
 
%\cite{Balantekin:2007es}
\bibitem{Balantekin:2007es}
  A.~B.~Balantekin, J.~Gava and C.~Volpe,
  %``Possible CP-Violation effects in core-collapse Supernovae,''
  Phys.\ Lett.\  B, {\bf 662}, 396 (2008)
  [arXiv:0710.3112 [astro-ph]].
  %%CITATION = PHLTA,B662,396;%%

\bibitem{2008arXiv0805.2717G} Gava, J. and Volpe, C., arXiv e-prints, arXiv:0805.2717 (2008)

\bibitem{Pantaleone:1992xh} J.~T.~Pantaleone, \prd, {\bf 46}, 510 (1992)

%\cite{Bronzan:1988wa}
\bibitem{Bronzan:1988wa}
  J.~B.~Bronzan,
  %``PARAMETRIZATION OF SU(3),''
  \prd, {\bf 38}, 1994 (1988).
  %%CITATION = PHRVA,D38,1994;%%

\bibitem{Betal80} V.~Barger, K.~Whisnant, S.~Pakvasa and R.~J.~N.~Phillips, \prd, {\bf 22}, 2718 (1980)

\bibitem{Naumov92} V.~A.~Naumov, Int. J. Mod. Phys. D, {\bf 1}, 379 (1992)

\bibitem{HS00} P.~F.~Harrison and W.~G.~Scott, Phys. Lett. B, {\bf 476}, 349 (2000)

\bibitem{T91} S.~Toshev, Mod. Phys. Lett. A, {\bf 6}, 455 (1991)

\bibitem{2002PhRvD..66g3005K} K.~Kimura, A.~Takamura and H.~Yokomakura \prd, {\bf 66}, 073005 (2002)

\bibitem{1997BrJPh..27..384B} J.~Bellandi, M.~M.~Guzzo and V.~M.~ Aquino, Brazilian Journal of Physics, {\bf 27}, 384 (1997)

\bibitem{K&M2009}  J.~P.~Kneller and G.~C.~McLaughlin, {\it in preparation}

\end{thebibliography}
\end{document}